\documentclass[11pt,tightenlines,eqsecnum,floats,aps,amsfonts,amsmath,amssymb,nofootinbib,prd,shownopacs,floatfix]{revtex4}
\usepackage[english]{babel}
\usepackage{amsfonts,amssymb,extarrows,mathrsfs}
\usepackage{xcolor}
\usepackage{physics}
\usepackage{empheq}
\usepackage{bbm}
\usepackage{pdfpages}
\usepackage{makecell}
\usepackage{tensor}
\usepackage{dsfont}
\usepackage{leftidx}
\usepackage{amsmath} 
\usepackage{amsfonts}
\usepackage{amsmath}
\usepackage{amssymb}
\usepackage{bbold}

\newcommand{\lr}{\left(}
\newcommand{\rr}{\right)}
\newcommand{\ls}{\left[}
\newcommand{\rs}{\right]}

\newcommand{\NES}{|n\rangle} 

\newcommand{\hp}{\hat{p}}
\newcommand{\ha}{\hat{a}}
\newcommand{\had}{\hat{a}^{\dagger}}

\newcommand{\hn}{\hat{n}}
\newcommand{\hH}{\hat{H}}
\newcommand{\uop}{\hat{\mathbbm{1}}} 

\newcommand{\sgn}{\text{sgn}}

\renewcommand{\d}{\mathrm{d}}
\newcommand{\kchf}[1]{\ensuremath{{}_1\text{F}_1\!\left(#1\right)}} 
\begin{document}
\title{Coherent states for fractional powers of the harmonic oscillator Hamiltonian}

\author{Kristina Giesel}
\email{kristina.giesel@gravity.fau.de}
\author{Almut Vetter}
\email{almut.vetter@fau.de}
\affiliation{Institute for Quantum Gravity, Theoretical Physics III,   Department of Physics, FAU Erlangen-N\"urnberg, \\ Staudtstr. 7, 91058 Erlangen, Germany}

\begin{abstract}
Inspired by special and general relativistic systems that can have Hamiltonians involving square roots, or more general fractional powers, in this article we address the question how a suitable set of coherent states for such systems can be obtained. This becomes a relevant topic if the semiclassical sector of a given quantum theory wants to be analysed. As a simple setup we consider the toy model of a deparametrised system with one constraint that involves a fractional power of the harmonic oscillator Hamiltonian operator and we discuss two approaches for finding suitable coherent states for this system.  In the first approach we consider Dirac quantisation and group averaging that have been used by Ashtekar et. al. but only for integer powers of operators. Our generalisation to fractional powers yields in the case of the toy model a suitable set of coherent states. The second approach is inspired by coherent states based on a fractional Poisson distribution introduced by Laskin, which however turn out not to satisfy all properties to yield good semiclassical results for the operators considered here and in particular do not satisfy a resolution of identity as claimed. Therefore, we present a generalisation of the standard harmonic oscillator coherent states to states involving fractional labels, which approximate the fractional operators in our toy model semiclassically more accurately and satisfy a resolution of identity. In addition, motivated by the way the proof of the resolution of identity is performed, we consider these kind of coherent states also for the polymerised harmonic oscillator and discuss their semiclassical properties.  
\end{abstract}

\maketitle
\section{Introduction}
\label{sectionIntroduction}
Semiclassical or more precisely coherent states were first introduced by Schr\"odinger \cite{schroedinger1926} to reassemble the classical equations of motion for the harmonic oscillator from its quantised version.
They moved back in the focus in optical applications due to work by Klauder and Glauber \cite{Glauber:1966,Klauder:1963} that includes for instance a generalisation  to multiple photon systems. Further generalisations for physical systems described by Lie groups were for example given in \cite{Perelomov:1972,Barut:1970,Perelomov:1986,Rasetti:1975} and a short instructive introduction can be found in \cite{Zhang+Feng+Gilmore:1990}. A classic compendium is the book \cite{Klauder+Skagerstam:1985} that summarises some aspects of coherent states and contains reprints of selected articles from the field of coherent states.
A modern comprehensive compendium on coherent states is the book \cite{Ali+Antoine+Gazeau:2014}.
Applications where one wants to have good control on the semiclassical sector are for example quantum cosmological models in case that one wants to make contact to classical cosmological models. In the context of loop quantum gravity (LQG) coherent states have also been used to test whether operators that are relevant in LQG have the correct semiclassical limit see for instance \cite{GCS2,GCS3,Sahlmann:2002qk,Brunnemann:2005in, Giesel:2006uk,Engle:2012yg,Han:2020chr}. 
The states will be suitable and will be called semiclassical states with respect to a certain operator, if the expectation value of the operator agrees with its corresponding classical observable up to small corrections. Semiclassical states are denoted as coherent states in case that they satisfy some additional properties which always includes the resolution of identity and depending on the specific application can include other properties like for instance being an eigenstate of the annihilation operator, see for example the discussion in \cite{Klauder:2001ra}.

In most non-relativistic models the Hamiltonian at the classical as well as at the quantum level is a polynomial in the elementary phase space variables or corresponding operators respectively. In a special or general relativistic framework there also exist operators in the corresponding quantum theory that involve square roots of the elementary operators associated with configuration and momentum variables, see for instance \cite{Laemmerzahl:1993xe,Brown:1994py,Kuchar:1995xn,Bicak:1997bx,Ashtekar:1997fb,Rovelli:1994ge,Domagala:2010bm,Giesel:2016gxq,gt2010,Giesel:2012rb,Alesci:2017yif}. The quantisation of these kind of operators often involves some ambiguities, therefore to analyse whether these operators yield the correct semiclassical limit can be understood as a basic consistency check any of these models needs to pass. This will in particular be of importance, if the spectrum of these operators is unknown which is often the case in the context of quantum gravity. In order to analyse the semiclassical limit, one wants to calculate the semiclassical limit of the physical operator which is the expectation value of this operator in suitable semiclassical states.  Therefore, if we are interested in such a semiclassical analysis we need to have semiclassical states that are suitable for approximating square root operators or more general operators with fractional powers well, that means yielding in leading order in $\hbar$ the corresponding classical expression. In the context of LQG there exist former work on semiclassical expectation values of the volume operator in \cite{Giesel:2006um}, which extends part of the techniques introduced in  the seminal papers \cite{GCS1,GCS2,GCS3,GCS4} from operators with integer powers to those with fractional powers. The semiclassical perturbation theory introduced in \cite{Giesel:2006um} was for instance used in \cite{Dapor:2017rwv,Han:2019feb} to obtain quantum cosmological models from full LQG and for instance in \cite{Giesel:2006uk,Han:2020chr} to analyse the semiclassical limit of LQG in a framework where the dynamics is encoded in a master constraint. In the context of quantum mechanics on a circle a semiclassical analysis of fractional operators has been recently considered in \cite{Giesel:2020jkz} where a slightly different technique using Kummer functions is considered. 
Although the techniques in \cite{Giesel:2006um,Giesel:2020jkz} might in principle be applicable to more general operators  as far as the fractional power is considered both works focus on operators that are either functions of only momentum or only configuration variables in their detailed analysis. In this article we want to address the question of suitable coherent states for Hamiltonian operators that are fractional powers of operator valued functions that involve momentum operators as well as operators associated to the configuration degrees of freedom. Such kind of coherent states will be needed in order to analyse the semiclassical limit of models like in \cite{Ashtekar:1997fb,Rovelli:1994ge,Domagala:2010bm,Giesel:2016gxq, Giesel:2012rb,Alesci:2017yif}. The complexifier coherent states introduced and analysed in \cite{GCS1,GCS2,GCS3,GCS4} are based on the strategy that one generalises the usual harmonic oscillator coherent states to the LQG framework. As far as the fractional operators in \cite{Giesel:2006um,Giesel:2020jkz} are considered one computes semiclassical expectation values still with respect to coherent states that are adapted to the dynamics of the harmonic oscillator and in the case of \cite{Giesel:2006um} in addition replaces the fractional operators by an operator valued power expansion.  Another strategy can also be to consider a set of coherent states that is more adapted to fractional operators than the harmonic oscillator coherent states. 

As a first step in this direction in this article we want to construct and use coherent states to calculate the semiclassical limit for fractional powers of the harmonic oscillator Hamiltonian as a simple toy model for testing semiclassical techniques in this context. We follow two different routes in the article. First we take into account that fractional operators in the context of gravitational systems often follow from constrained systems, where the dynamics of the physical degrees of freedom is generated by Hamiltonians that involve fractional powers, see for instance \cite{Laemmerzahl:1993xe,Domagala:2010bm,gt2010,Giesel:2016gxq, Giesel:2012rb,Alesci:2017yif} for examples. Therefore, we consider for our quantum mechanical toy model a Dirac quantisation for the construction of physical coherent states along the lines of \cite{Ashtekar:2005dm}, where group averaging techniques introduced in \cite{KLAUDER1997419} are used to obtain physical coherent states adapted to the constraints. These states can then be used to compute semiclassical expectation values and we analyse whether the so obtained physical coherent states are well adapted to fractional powers of the harmonic oscillator Hamiltonian. As already pointed out in \cite{Ashtekar:2005dm}, compared to the kinematical coherent states the physical ones have some restriction on their labels and this will enter in a non-trivial way into the computation of semiclassical expectation values. The results in \cite{Ashworth:1996yv,Ashtekar:1997fb} address the case of constraint operators with integer powers, in particular only linear and quadratic polynomials in the elementary operators. In order to be able to apply their techniques also to constraints with fractional powers as a first step we apply a kind of Euler rescaling \cite{Struckmeier:2005} at the level of the constraints. In a second step we then combine techniques from \cite{Ashworth:1996yv,Ashtekar:1997fb} with results from \cite{Giesel:2020jkz} to perform a semiclassical analysis.  In a second approach inspired by the work in \cite{2003:Laskin1,2009:Laskin2} on coherent states based on a fractional Poisson distributions, we want to address the question whether these kind of states provide a set of good semiclassical states for fractional powers of the harmonic oscillator. Likewise to the formalism in \cite{Giesel:2006um} the fractional operator is also substituted by integer powers of the same operator. But since we also consider different coherent states, this kind of substitution takes a simpler form as in \cite{Giesel:2006um}. It turns out that the coherent states introduced in \cite{2009:Laskin2} do not provide a good semiclassical approximation for the fractional operators in our toy model and furthermore they do not satisfy a resolution of identity as claimed in \cite{2009:Laskin2}.  We introduce a set of coherent states which slightly differs from those states and show that this set has better semiclassical properties for the fractional operators under consideration in this toy model, as for instance satisfying a resolution of identity. We discuss such kind of coherent states also in the context for a polymere quantisation, see for instance \cite{Ashtekar:2002sn,BandStructurePolymer2013} of the fractional harmonic oscillator Hamiltonian and show that the results agree with the ones in the Schrödinger case up to corrections that involve the discretisation scale that enters the framework of the polymere formulation of quantum mechanics. The generalised coherent states considered here are different ones than the shadow states considered in \cite{Ashtekar:2002sn} and we analyse their semiclassical properties for our toy model.

The article is structured as follows: 
After the introduction in section \ref{sectionIntroduction},  in section \ref{sec:ConstCS} we apply the techniques of  \cite{Ashworth:1996yv,Ashtekar:2005dm}  to the case of deparametrised constrained systems with a fractional Hamiltonian. For this purpose we work in the extended phase space and consider a kind of Euler rescaling of the temporal coordinate which allows to shift the fractional power of the Hamiltonian to the momentum of the temporal coordinate. This step brings the constraint into a form for which the corresponding semiclassical expectation values can be computed in terms of Kummer functions as presented in \cite{Giesel:2020jkz} with a good semiclassical behaviour at the level of kinematical coherent states. Then we apply group averaging to construct physical coherent states and use those to compute semiclassical expectation values of fractional Hamiltonians in terms of integer power operators by taking into account the interplay between the fractional power of the temporal momentum and the Hamiltonian. Further, we discuss differences at the kinematical and physical level as well as in which sense these results yield a good approximation to the fractional power operator's semiclassical expectation values. Moreover, we compare within our toy model the results obtained from Dirac quantisation with the results one gets using reduced phase space quantisation. In section \ref{sec:GenCS} inspired by the coherent states based on a fractional Poisson distribution introduced in \cite{2009:Laskin2} we analyse how the standard harmonic oscillator coherent states can be better adapted to operators with fractional powers in a reduced phase space quantisation approach.
In subsection \ref{sec:ReviewLaskinStates} we briefly review the main results for the coherent states introduced by Laskin in order to explain in subsection  \ref{sectionGeneralisedCoherentStatesForFractionalHamiltonians} where the proof of the resolution of identity in \cite{2009:Laskin2} fails to work and generalise the construction of the usual harmonic oscillator coherent states to states involving fractional labels that satisfy a resolution of identity. Since this requires to extend the range of the angular coordinate as suggested in \cite{Klauder:1995yr}, we further discuss the construction of those kind of states also in the polymere framework of quantum mechanics, analyse their semiclassical properties and compare the results to the Schrödinger case in subsection \ref{sec:PolyHam}. We end this section in subsection \ref{sectionSemiclassPerturbationFractionalHamiltonians} with a comparison of the results one obtains with the  semiclassical perturbation theory from \cite{Giesel:2006um} in order to be able to compare our results obtained here to the already existing ones in the literature. In appendix \ref{sectionComplexifierCoherentStatesForFractionalHeatKernel} we  briefly discuss how such coherent states are related to complexifier coherent states associated with the standard and fractional heat equation.  In this work here we will only focus on fractional powers of the harmonic oscillator, since this provides a toy model where these techniques can be easily tested and compared to already existing results in the literature. We comment in the  conclusions about the possibility to generalise these techniques to more complicated systems which can be explored in future work.

\section{Coherent states in constrained systems: Dirac quantisation and application to fractional Hamiltonians}
\label{sec:ConstCS}
There exists already preliminary work on the construction of coherent states for constrained systems in the literature, like for instance in \cite{Ashtekar:2005dm, Ashworth:1996yv} where physical coherent states for constrained systems were constructed. In \cite{Ashworth:1996yv} the physical coherent states are deduced from the inclusion of constraints into the framework of quantum mechanical path integrals which results in projecting a state from the kinematical Hilbert space into a state in the physical Hilbert space. The article concludes with the application of this method to time reparametrisation invariant systems which for example occur in quantum gravity. The method in \cite{Ashtekar:2005dm} starts from known kinematical harmonic oscillator coherent states and projects them with the help of group averaging to the physical Hilbert space. If it is assumed that the coherent states are peaked on the classical constraint surface, the results in \cite{Ashtekar:2002sn} show that physical coherent states as well as their inner product can be obtained. For an application in cosmology to Bianchi I spacetimes, see \cite{Bolen:2004tq}. The work in \cite{Ashtekar:2005dm} considers constraints with an either linear or quadratic dependence on the elementary phase space variables only. In this section we want to follow closely the methods introduced in \cite{Ashtekar:2005dm} but now apply them to constraints that involve fractional powers of the elementary phase space variables.

\subsection{Euler rescaling as canonical transformations on the extended phase space}
\label{sectionEulerRescaling}
In order  to deal with a constrained system with a fractional power of a Hamiltonian we work in the extended phase space.  As discussed in \cite{Ashtekar:1997fb} coherent states for a constrained system will in general have some restriction on their label in order to ensure that their labels are consistent with the constraints of the system. As mentioned above we will restrict our discussion to fractional powers of the harmonic oscillator Hamiltonian here. However, the strategy can be carried over to more complicated systems, if the constraint associated with temporal diffeomorphisms (for general relativity this is the Hamiltonian constraint)  can be written in deparametrised\footnote{Deparametrisation in this context means that the constraint can be written as $C=p_T+h$, where $p_T$ denotes the momentum of the configuration variable playing the role of the clock of the system and h involves only the remaining phase space variables but not the one from the clock degrees of freedom.} form at the classical level and the set of coherent states that one wants to use for the computations have good semiclassical properties as far as integer powers of the Hamiltonian are considered. We will discuss this aspect in more detail in our conclusions in section \ref{sec:SumConcl}. To explain how the Euler rescaling can be useful in this context, let us consider the following set up: we examine a Hamiltonian that is given by some fractional power of the harmonic oscillator in one dimension formulated on the phase space $T^*Q$ with elementary variables $(q,p)$. We denote the Hamiltonian as $H_{\rm HO}^\mu$ where $\mu$ is a rational number $\mu=\frac{v}{w}$ with $v,w\in\mathbb{N}$ and $H_{\rm HO}$ is the Hamiltonian of the harmonic oscillator that is given by $H_{\rm HO}=\frac{p^2}{2m}+\frac{m\omega_0^2q^2}{2}$. In order to map this dynamical system into a constrained system with a deparametrised constraint we work in  the extended phase space $T^*M$ in which the temporal coordinate is also treated as a canonical variable with coordinates $(t,p_t,q,p)$. The constraint of the system in the extended phase space has the form
\begin{equation*}
C=k(p_t+H_{\rm HO}^\mu), \quad \{q,p\}=1,\quad \{t,p_t\}=1,
\end{equation*}
where $k$ is some arbitrary real and non-zero number and all remaining Poisson brackets vanish. Let us briefly comment on the units of the involved quantities. From the constraint $C$ we can read off that 
$[p_t]=[{\rm energy}]^\mu=J^\mu$. Furthermore we have $[q]=[{\rm length}]=m, [p]=[force\times time]=Ns$ and $[t]=J^{-\mu+1}s$. Here deparametrisation means that the constraint can be written linearly in the temporal momentum and the remaining part of the constraint does not include $t$. In the extended phase space as shown for instance in \cite{Struckmeier:2005}, we can write down a set of first order Hamilton's equations with respect to an evolution parameter that we denote by $s$
\begin{equation}
\label{eq:EOMextPS}
\frac{dq(s)}{ds}=k\{q(s),H_{\rm HO}^\mu\},\quad  \frac{dp(s)}{ds}=k\{p,H_{\rm HO}^\mu\},\quad
\frac{dt}{ds}=\{t,C\}=k,\quad \frac{dp_t(s)}{ds}=\{p_t,C\}=0.
\end{equation}
Now what we are interested in a constrained system is the dynamics of the observables, which are phase space functions in the so-called reduced phase space. The reduced phase space can be obtained by a symplectic reduction with respect to $C$ and can be coordinatised by the corresponding elementary observables associated with $q,p$. These observables are quantities that are required to commute with the constraint $C$. From now on let us consider the choice $k=1$. In this case 
the physical Hamiltonian, which generates the evolution of the observables, is then given by the function $H_{\rm HO}^\mu$ evaluated at the observables of $q$ and $p$. Let us denote the observables of $q,p$ by $O_q$ and $O_p$, then the classical Hamilton's equation in the reduced phase space read:
\begin{equation}
\label{eq:EOMextPS2}
\frac{dO_q}{d\tau}=\{O_q(\tau),H_{\rm HO}^\mu(O_q,O_p)\},\quad  \frac{dO_p(\tau)}{d\tau}=\{O_p,H_{\rm HO}^\mu(O_q,O_p)\},\quad
\end{equation}
where we denoted the evolution parameter in the reduced phase space by $\tau$ to match with our later notation at the end of section \ref{sec:ConstCS}. We realise that for the choice $k=1$ and under the identification $O_q\to q$, $O_p\to p$ and $\tau\to s$ the Hamilton's equations in \eqref{eq:EOMextPS} and \eqref{eq:EOMextPS2} agree for this subset of variables. In this sense we can cast any classical Hamiltonian system with a given Hamiltonian $H$ into a constrained system with constraint $C=p_t+H$ in the extended phase space that is written linearly in the temporal momentum. Looking at the equations of motion in (\ref{eq:EOMextPS}) with $k=1$, we realise that the outer derivative of $H_{\rm HO}^\mu$ involved the first order equations for $\frac{dq}{ds}=\mu H_{\rm HO}^{\mu-1}\{q,H_{\rm HO}\}$ and likewise for $\frac{dp}{dt}$ can be absorbed into a redefinition of the temporal coordinate and with respect to the transformed time the Hamiltonian is just linear in the harmonic oscillator Hamiltonian $H_{\rm HO}$. In the extended phase space this can be formulated as a canonical transformation of the form
\begin{equation}
\label{eq:CanTrafo}
P_T= |p_t|^\frac{1}{\mu}\sgn(p_t),\quad T=\sgn(p_t)\frac{\mu t}{|p_t|^{\frac{1}{\mu}-1}},\quad Q=q,\quad P=p.  
\end{equation}
The variables $(T,P_t)$ have the units $[P_T]=J$ and $[T]=s$. This transformation often denoted as Euler rescaling was discussed in a more general context for instance in \cite{Struckmeier:2005}. Note that in our case this is rather a kind of dual Euler rescaling, since here the new temporal momentum $P_T$ is a function of $p_t$ only, whereas the new temporal coordinate $T$ is a function of $t,p_t$. In contrast to the Euler rescaling in \cite{Struckmeier:2005} the new temporal variable  $T$ is a function of $t$ only and $P_T$ a function of $t,p_t$. Furthermore in \cite{Struckmeier:2005} the transformation to $t$ involves an integral. Assuming that $p_t\not=0$ we can multiply the entire constraint $C$ by $|p_t|^{\frac{1}{\mu}-1}\sgn(p_t)$ and obtain
\begin{eqnarray}
\label{eq:EquivConstr}
\widetilde{C}&=&|p_t|^{\frac{1}{\mu}}+|p_t|^{\frac{1}{\mu}-1}\sgn(p_t)H^{\mu}_{\rm HO}\approx 
|p_t|^{\frac{1}{\mu}}-H_{\rm HO} H_{\rm HO}^{-\mu}\sgn(H_{\rm HO})H_{\rm HO}^\mu 
=|p_t|^{\frac{1}{\mu}} - H_{\rm HO},
\end{eqnarray}
where we used  the weak $\approx$ equivalence of quantities on the constraint surface $C=0$ and that $\sgn(H_{\rm HO})=1$ since $H_{\rm HO} >0$. In this sense the new constraint $\widetilde{C}$ implies $p_t=-H_{\rm HO}^\mu$ on the constraint surface which requires $p_t<0$ and thus $-p_t=|p_t|\approx H_{\rm HO}^\mu$ leading to $|p_t|^\frac{1}{\mu}\approx H_{\rm HO}$.  An important property of the above defined canonical transformation is that $p_t$, and thus also any function of it, is a constant of motion which on the reduced phase space can be identified with the energy of the physical system. As a consequence, when we use the rewritten and equivalent version of the constraint in (\ref{eq:EquivConstr}) in the next subsection to construct coherent states in constrained systems, we have to take this into account and consider that not $H_{\rm HO}$ is the energy of our original system that we start from but $H^{\mu}_{\rm HO}$ and thus $ H_{\rm HO}=|p_t|^{\frac{1}{\mu}}=(E^{(s)})^{\frac{1}{\mu}}$, where $E^{(s)}$ denotes the energy of the system and can be determined once the phase space variables are given. To keep track of the original definition of the energy of the system before the dual Euler rescaling has been applied goes in the same direction as the idea of a kind of reference metric suggested by Klauder in \cite{Klauder:1997} in order to be able to have a consistent interpretation of the dynamical operator even if a transformation of the phase space variables has been applied. 

If we want to work with the constraint $\widetilde{C}$ in (\ref{eq:EquivConstr}), then it will look like that we have not gained much, since we just moved the fractional power from the Hamiltonian to the momentum $p_t$. However, as shown in \cite{Giesel:2020jkz} using Kummer functions fractional powers of the momentum operator can be well approximated by the standard harmonic oscillator coherent states and we will use those results here to obtain appropriate coherent states on the kinematical Hilbert space which approximate the quantum constraint well semiclassically. 

Now given the constraint in the form we wanted, we can proceed in two directions. Either we consider Dirac quantisation and solve the constraint in the quantum theory or we derive the reduced phase at the classical level and apply reduced phase space quantisation. At this stage both are equally justified. In the context of coherent states this carries over to the situation that when applying Dirac quantisation those coherent states are usually constructed on the kinematical Hilbert space. However, in order to actual compute relevant semiclassical expectation values one would like to use physical coherent states that encode some information about the constraints in the system. As mentioned at the beginning of this section a strategy to obtain physical coherent states from a given set of kinematical coherent states was presented in \cite{Ashtekar:2005dm} and applied to a couple of examples there. We will follow this strategy in the next subsection and apply it to fractional Hamiltonians combined with the Euler rescaling just discussed, where this technique is still based on using the standard harmonic oscillator coherent states. At the end of the next subsection We will also show that in this case reduced phase space quantisation and Dirac quantisation will yield to the same set of physical coherent states. 
.

\subsection{Physical Coherent states for constraints with fractional Hamiltonians}
\label{sectionPhysicalCoherentStatesForFractionalHamiltonians}
We want to apply the techniques introduced in \cite{Ashtekar:2005dm} to our fractional powers $\mu$ of the harmonic oscillator Hamiltonian which we shortly refer to as fractional Hamiltonians.
Instead of considering the fractional Hamiltonians directly, we  go over to the extended phase space as described in section \ref{sectionEulerRescaling} and consider a constraint of the form
\begin{align}
 C = p_t + H_{\rm{HO}}^{\mu} = 0,
\end{align}
where $p_t$ is the canonical conjugate momentum to a new time variable $t$
and a constant of motion with respect to the fractional Hamiltonian in consideration.
For fixed phase space coordinates $q,p$ the temporal momentum $p_t$ corresponds to the negative
energy of the system, that is $p_t = -E^{(s)}$ with $E^{(s)} > 0$.
Notice that for $\mu=1$ this reduces to the case for the harmonic oscillator.
Because of the general fractional power $\mu$ of the harmonic oscillator Hamiltonian the constraint
in general might be difficult to handle. Therefore, we transform the constraint using the dual Euler rescaling to obtain an equivalent constraint as displayed in (\ref{eq:EquivConstr}) in section \ref{sectionEulerRescaling} which reads
\begin{align}
\widetilde{C} =|p_t|^{\frac{1}{\mu}}- H_{\rm{HO}} 
     = |p_t|^{\frac{1}{\mu}}-\left(\frac{p^2}{2m} + \frac{1}{2} m \omega_0^2 q^2\right) 
     = |p_t|^{\frac{1}{\mu}}-\hbar \omega_0 \bar{z} z \approx 0
\end{align}
for $z=\sqrt{\frac{m\omega_0}{2\hbar}}q+i\sqrt{\frac{1}{2\hbar m\omega_0}}p \in \mathbb{C}$. The kinematical Hilbert space of this model is ${\cal H}_{\rm kin}={\cal H}_1\otimes {\cal H}_2=L_2(\mathbb{R},{\rm d}q)\otimes L_2(\mathbb{R},{\rm dp}_t)$, where we use for both Hilbert spaces the standard Schr\"odinger representation, i.e.\,for the first one the occupation number representation and for the second one the momentum representation. The kinematical inner product for two kinematical states $|\Psi\rangle=|\psi_1\rangle\otimes|\psi_2\rangle$ and $|\Psi'\rangle=|\psi_1\rangle'\otimes|\psi_2'\rangle$ has the following form
\begin{equation}
\langle \Psi\, |\, \Psi'\rangle_{\rm kin}=\langle \psi_1\, |\, \psi_1'\rangle_{{\cal H}_1} \langle \psi_2\, |\, \psi_2'\rangle_{{\cal H}_2}.   
\end{equation}
The constraint operator is then just given by
\begin{align*}
\hat{\widetilde{C}}=\uop_{{\cal  H}_1}\otimes|\hat{p}_t|^{\frac{1}{\mu}} \uop_{{\cal  H}_2}-\hbar\omega_0\lr\hat{a}^\dagger\hat{a}+\frac{1}{2}\rr\uop_{{\cal  H}_1}\otimes\uop_{{\cal H}_2}.
\end{align*}
As a first step we define kinematical coherent states whose expectation value of $\hat{\widetilde{C}}$ reproduces to lowest order in $\hbar$ the classical constraint. These kinematical coherent states can be obtained from a tensor product of the standard harmonic oscillator coherent states as follows
\begin{equation}
|\Psi_{\alpha,(t^0,p_t^0)}\rangle:=|\Psi_\alpha\rangle\otimes |\Psi_{t^0,p_t^0}\rangle,     
\end{equation}
where $\alpha:=\sqrt{\frac{m\omega_0}{2\hbar}}q_0+i\sqrt{\frac{1}{2\hbar m\omega_0}}p_0$ and $(t^0,p_t^0)$ are classical labels associated with the extended phase space. The explicit form of these states is given by
\begin{align}
|\Psi_\alpha\rangle &= e^{-\frac{|\alpha|^2}{2}}\sum\limits_{n=0}^\infty \frac{\alpha^n}{\sqrt{n}}|n\rangle
\end{align}
and
\begin{align}
|\Psi_{t^0,p_t^0}\rangle &=\int\limits_{\mathbb{R}}{\rm d}p_t \Psi_{t^0,p_t^0}(p_t)|p_t\rangle = \int\limits_{\mathbb{R}}{\rm d}p_t C_{t^0,p_t^0,\hbar}e^{-\frac{(p_t-p_t^0)^2}{2((\hbar\sigma)^\mu)^2}}e^{-\frac{i}{\hbar} p_t t^0}|p_t\rangle 
\end{align}
with $\sigma$ carrying units $[\sigma]=s^{-1}$ such that the arguments of all exponentials are dimensionless. If we define a similar dimensionless label $\alpha_t$ also for the temporal phase space coordinates, then $\sigma$ will enter as $\alpha_t:=\frac{\hbar}{\sqrt{2}(\hbar\sigma)^\mu}(\frac{(\hbar\sigma)^{2\mu}}{\hbar^2}t+\frac{i}{\hbar}p_t)$.  The coherent state $|\Psi_\alpha\rangle$ is already normalised and we choose $C_{t^0,p_t^0,\hbar}=\frac{1}{(\pi^{\frac{1}{4}}\hbar\sigma)^\mu}e^{\frac{i}{\hbar}p_t^0t^0}
e^{-\frac{(p_t^0)^2}{2(\hbar\sigma)^{2\mu}}}$ such that also $|\Psi_{t^0,p_t^0}\rangle$ is normalised and thus $|\Psi_{\alpha,(t^0,p_t^0)}\rangle$ as well. The semiclassical expectation value of the constraint operator $\hat{\widetilde{C}}$ can be computed as 
\begin{align}
\langle\Psi_{\alpha,(t^0,p_t^0)}|\,\hat{\widetilde{C}} \,|\Psi_{\alpha,(t^0,p_t^0)}\rangle &=
-\langle\Psi_\alpha|\,\hbar\omega_0(\hat{a}^\dagger\hat{a}+\frac{1}{2})\,|\Psi_\alpha\rangle
+
\langle\Psi_{t^0,p_t^0}|\,|\hat{p}_t|^{\frac{1}{\mu}}\,|\Psi_{t^0,p_t^0}\rangle \\ \nonumber
&=-\hbar\omega_0(\overline{\alpha}\alpha+\frac{1}{2})
+\langle\Psi_{t^0,p_t^0}|\,|\hat{p}_t|^{\frac{1}{\mu}}\,|\Psi_{t^0,p_t^0}\rangle.
\end{align}
Using the techniques presented in \cite{Giesel:2020jkz}, we can express the second semiclassical expectation value in terms of Kummer functions and obtain
\begin{align}
 \langle\Psi_{\alpha,(t^0,p_t^0)}|\,\hat{\widetilde{C}} \,|\Psi_{\alpha,(t^0,p_t^0)}\rangle &=
-\frac{p_0^2}{2m}-\frac{m\omega^2_0q_0^2}{2}-\frac{\hbar\omega_0}{2}
+\frac{\Gamma(\frac{\frac{1}{\mu}+1}{2})}{\sqrt{\pi}}
\left((\hbar\sigma)^\mu\right)^{\frac{1}{\mu}}\kchf{-\frac{1}{2\mu},\frac{1}{2},-\frac{(p^0_t)^2}{(\hbar\sigma)^{2\mu}}},
\end{align}
here $\kchf{a,b,z}$ with $z\in\mathbb{C}$ denotes the Kummer function of the first kind also called the  confluent hypergeometric function of the first kind. For more details on Kummer functions and particularly on how their Fourier transform can be used to obtain the above semiclassical expectation value we refer the reader to the work in \cite{Giesel:2020jkz}. 
As far as the semiclassical computations are concerned we are interested the sector where $\hbar$ is small compared to one, which allows us to express the semiclassical expectation value as an expansion in (fractional) powers of $\hbar$. The classical limit can then be obtained in the limit where we send $\hbar\to 0$. Consequently in the case of the Kummer function we can use its asymptotic behaviour for large arguments which is well known. As shown in \cite{Giesel:2020jkz}, the relevant asymptotic expansion for the semiclassical expectation value is given by
\begin{equation}
\langle\Psi_{t^0,p_t^0}|\,|\hat{p}_t|^{\frac{1}{\mu}}\,|\Psi_{t^0,p_t^0}\rangle
\approx 
|p^0_t|^{\frac{1}{\mu}}\sum\limits_{n=0}^\infty
\frac{(-\frac{1}{2\mu})_n(\frac{\mu-1}{2\mu})_n}{n!}\left(\frac{(\hbar\sigma)^{2\mu}}{(p^0_t)^2}\right)^n,
\end{equation}
where $(a)_n$ denotes the Pochhammer symbols also called raising factorials with $(a)_0=1, (a)_1=a$ and $(a)_n=a(a+1)\cdots(a+n-1)$. Given these asymptotics of the Kummer function we obtain for the semiclassical expectation value of $\hat{\widetilde{C}}$
\begin{align}
\label{eq:KummExp}
\langle\Psi_{\alpha,(t^0,p_t^0)}|\,\hat{\widetilde{C}} \,|\Psi_{\alpha,(t^0,p_t^0)}\rangle\approx
-\frac{p_0^2}{2m}-\frac{m\omega^2q_0^2}{2}-\frac{\hbar\omega_0}{2}
+
|p^0_t|^{\frac{1}{\mu}}\left(1-\frac{\frac{1}{\mu}(1-\frac{1}{\mu})}{4}\frac{(\hbar\sigma)^{2\mu}}{(p^0_t)^2}+o(\hbar^{4\mu})\right) \nonumber \\
=
|p^0_t|^{\frac{1}{\mu}}-H_{\rm HO}+\hbar\frac{\omega_0}{2}-|p^0_t|^{\frac{1}{\mu}}\frac{\frac{1}{\mu}(1-\frac{1}{\mu})}{4}\frac{(\hbar\sigma)^{2\mu}}{(p^0_t)^2}+o(\hbar^{4\mu}),
\end{align}
where used $H_{\rm HO}=\frac{p_0^2}{2m}+\frac{m\omega^2_0q_0^2}{2}$. Hence, in the semiclassical limit $\hbar\to 0$ we recover the classical constraint $\widetilde{C}$
\begin{equation}
\lim\limits_{\hbar\to 0} \langle\Psi_{\alpha,(t^0,p_t^0)}|\,\hat{\widetilde{C}} \,|\Psi_{\alpha,(t^0,p_t^0)}\rangle
=|p^0_t|^{\frac{1}{\mu}}-H_{\rm HO}=\widetilde{C}.
\end{equation}
The point that we obtain in the limit $\hbar\to 0$ the correct classical expression confirms the theorem in \cite{GCS3} based on the Hamburger momentum problem by explicit computations in our toy model\footnote{Note that there exist classical labels of the coherent states for which the corresponding semiclassical expectation values might not satisfy the assumptions of the theorem.}. Due to the fact that using the techniques introduced in \cite{Giesel:2020jkz}, we can also explicitly compute the higher than leading order terms. In this sense our results extend those in \cite{GCS3} concerning the formalism for non-polynomials operators. Note that for the special case that $\mu=\frac{1}{2n}$ with $n\in\mathbb{N}$ we have $\frac{1}{\mu}=2n$ and then the first argument of the Kummer function is $-n$ and in this case it can be expressed in terms of Hermite polynomials yielding for instance the expected semiclassical expectation value for $p_t^2$ for the choice of $n=1$. The rather unusual powers of $\hbar$ involving $\mu$ are due to the fact that in our case the unit of $p_t$ is $[p_t]=J^\mu$, whereas for the spatial coordinates one uses the characteristic length of the harmonic oscillator $\ell:=\sqrt{\frac{\hbar}{m\omega}}$ to introduce dimensionless quantities and $\ell^2$ is linearly in $\hbar$. For odd integers we have that $p_t^n$ can also become negative but then even at the classical level due to the fact that $H_{\rm HO}>0$ the constraint $p_t^\frac{1}{\mu}-H_{\rm HO}\approx 0$ has no solutions and that is why we work with $|p_t|$ here. Note that this is similar to the situation for the reference matter models where one usually also restricts to certain parts of the full phase space by restricting the sign of the clock momentum, see for instance the discussion in \cite{Giesel:2007wi,Domagala:2010bm,Giesel:2012rb,Husain:2012,Giesel:2016gxq,ghls2020,Giesel:2020jkz}.

The discussion so far was completely at the kinematical level, therefore we will apply the group averaging procedure to obtain physical coherent states along the lines of \cite{Ashtekar:2005dm}. In our case the group averaging operator is given by
\begin{align}
\hat{U}(\lambda)
&=e^{-\frac{i\lambda}{\hbar\omega_0}\left(\uop_{{\cal  H}_1} \otimes|\hp_t|^\frac{1}{\mu} \uop_{{\cal  H}_2}-\hbar\omega_0(\had \ha +\frac{1}{2})\uop_{{\cal  H}_1} \otimes \uop_{{\cal  H}_2} \right)}
=e^{i\lambda(\hn +\frac{1}{2} \uop_{{\cal  H}_1})}\otimes e^{-\frac{i\lambda}{\hbar\omega_0}|\hp_t|^{\frac{1}{\mu}}\uop_{{\cal  H}_2}} \, , 
\end{align}
where we used that $\hp_t$ commutes with $\hH_{\rm HO}$, rewrote $\hH_{\rm HO}$ in terms of the number operator $\hn=\had \ha$ and rescaled the constraint by $\hbar\omega_0$ in order to obtain a dimensionless quantity. Now we calculate the action of the unitary operator involved in the group averaging $\hat{U}(\lambda) = e^{- i\lambda \hat{\widetilde{C}}}$
on the kinematical coherent state $\Psi_{\alpha,(t_0,p_t^0)}$  leading to
\begin{align}
 e^{- \frac{i\lambda}{\hbar\omega_0} \hat{\widetilde{C}}} | \Psi_{\alpha,(t^0,p_t^0)} \rangle 
 &= e^{i\lambda (\hn+\frac{1}{2}\uop_{{\cal  H}_1})} e^{-\frac{|\alpha|^2}{2}}
   \sum\limits_{n=0}^{\infty}  \frac{\alpha^n}{\sqrt{n!}}\NES \otimes e^{-\frac{i\lambda}{\hbar\omega_0} |\hat{p}_t|^{\frac{1}{\mu}}\uop_{{\cal  H}_2}}|\Psi_{t^0,p_t^0}\rangle\nonumber \\
    &=
   e^{-\frac{|\alpha|^2}{2}}
     \int\limits_{\mathbb{R}}\,{\rm d}p_t\sum\limits_{n=0}^{\infty} e^{i\lambda (n+\frac{1}{2})}  \frac{\alpha^n}{\sqrt{n!}}\NES   
    \otimes e^{-\frac{i\lambda}{\hbar\omega_0}|p_t|^{\frac{1}{\mu}}}\Psi_{t^0,p_t^0}(p_t) |p_t\rangle  \nonumber \\
    &=
       e^{-\frac{|\alpha|^2}{2}} \int\limits_{\mathbb{R}}\,{\rm d}p_t\sum\limits_{n=0}^{\infty}\,
       e^{-\frac{i\lambda}{\hbar\omega_0}\big(|p_t|^{\frac{1}{\mu}}- \hbar\omega_0(n+\frac{1}{2})\big)}  \frac{\alpha^n}{\sqrt{n!}}\Psi_{t^0,p_t^0}(p_t)\NES 
    \otimes |p_t\rangle ,
\end{align}
where $\Psi_{t^0,p_t^0}(p_t)$ denotes, as before, the standard coherent state in the momentum representation. Next we apply the group averaging to obtain physical coherent states which in our case will not be elements of ${\cal H}_{\rm kin}$ but distribution on a dense subset ${\cal S}\subset{\cal H}_{\rm kin}$, following  closely the formalism in \cite{Ashtekar:2005dm}. In addition we introduce a projection operator $\widehat{P}_{p_t<0}$ that projects on the negative part of the spectrum of $\hat{p}_t$ to ensure that the classical condition $p_t=-H^{\mu}_{\rm HO}$ which requires $p_t<0$ is also fulfilled at the quantum level. This projection operator can be implemented via $\widehat{P}_{p_t<0}:=\mathds{1}_{{\cal H}_1}\otimes\theta(-\hat{p}_t)$, where $\theta$ denotes the usual Heaviside function that vanishes if $p_t\geq 0$. Then we  obtain the physical constrained coherent states as follows
\begin{align}
\label{eq:NonNormalisedPhysicalCS}
 |\Psi_{\alpha,(t^0,p_t^0)}^{\rm{phy}}\rangle 
 &= \frac{1}{K} \widehat{P}_{p_t<0}\int\limits_{\mathbb{R}} \mathrm{d}\lambda \, \hat{U}(\lambda) 
    |\Psi_{\alpha,t^0,p_t^0)}\rangle 
 =   \frac{1}{K}\int\limits_{\mathbb{R}} \mathrm{d}\lambda \,  \widehat{P}_{p_t<0}\hat{U}(\lambda) 
    |\Psi_{\alpha,t^0,p_t^0)}\rangle 
    \nonumber \\
 &= \frac{e^{-\frac{|\alpha|^2}{2}}}{K} \int\limits_{\mathbb{R}} \mathrm{d}\lambda  \int\limits_{\mathbb{R}} \mathrm{d}p_t  \sum\limits_{n=0}^{\infty} \,\theta(-p_t)e^{-\frac{i\lambda}{\hbar\omega_0}\big(|p_t|^{\frac{1}{\mu}}- \hbar\omega_0(n+\frac{1}{2})\big)}\frac{\alpha^n}{\sqrt{n!}}\Psi_{t^0,p_t^0}(p_t)\NES\otimes |p_t\rangle  \nonumber \\
 &= \frac{2\pi e^{-\frac{|\alpha|^2}{2}}}{K}\int\limits_{\mathbb{R}} \mathrm{d}p_t  \sum\limits_{n=0}^{\infty}\theta(-p_t)\delta\Big(\frac{|p_t|^{\frac{1}{\mu}}}{\hbar\omega_0}-(n+\frac{1}{2})\Big)\frac{\alpha^n}{\sqrt{n!}}\Psi_{t^0,p_t^0}(p_t)\NES \, \otimes |p_t\rangle \nonumber \\
 &=\frac{2\pi e^{-\frac{|\alpha|^2}{2}}}{K}\int\limits_{\mathbb{R}} \mathrm{d}p_t \sum\limits_{n=0}^{\infty}\hbar\omega_0\mu(\epsilon_n)^{\mu-1}\delta(p_t+\epsilon_n^\mu)\frac{\alpha^n}{\sqrt{n!}}\Psi_{t^0,p_t^0}(p_t)\NES \, \otimes |p_t\rangle  ,
\end{align}
where we interchanged the order of the integration  over $\lambda$ with the summation and integration over $p_t$, used the definition of the Fourier transform of the delta function and defined $\epsilon_n:=\hbar\omega_0(n+\frac{1}{2}),\, n\in\mathbb{N}_0$ in the last step.
Here $K$ is a real constant whose value can be chosen such that the resulting physical coherent states are normalised as done in (\ref{eq:normalisedPhysicalConstarinedCS}) below.
Because the spectrum of $\hat{p}_t$ is the entire real line we have that even if we project to its negative part, that ${\rm spec}(|\hat{p}_t|^{\frac{1}{\mu}})\cap{\rm spec}(\hat{H}_{\rm HO})\not=\emptyset$ and thus one obtains a non-trivial distribution after group averaging. Similarly to the example of the linear constraint in \cite{Ashtekar:2005dm}, where also a distributional physical coherent state is obtained, the result of the group averaging can be understood as the restriction of the kinematical coherent state to the constraint surface with an additional modification in the measure. The physical inner product can be explicitly computed and reads
\begin{align}
\langle\,\Psi^{\rm phy}_{\alpha,(t^0,p_t)}|\, \Psi^{\rm phy}_{\beta, ({t'}^{0},{p'}^0_{t})}\rangle 
&=
\frac{1}{K}\int\limits_{\mathbb{R}}{\rm d}\lambda\, \langle \widehat{P}_{p_t<0}\hat{U}(\lambda)\Psi_{\alpha,(t^0,p_t^0)}\,|\, \Psi_{\beta,({t'}^0,{p'}_t^0)}\rangle \nonumber \\
&=
\frac{2\pi\mu\hbar\omega_0}{K}e^{-\frac{|\alpha|^2+|\beta|^2}{2}}\sum\limits_{n=0}^{\infty}\frac{(\overline{\alpha}\beta)^n}{n!}\epsilon_n^{\mu-1}\overline{\Psi}_{t^0,p_t^0}(-\epsilon^\mu_n)\Psi_{{t'}^0,{p'}_t^0}(-\epsilon^\mu_n).
\end{align}
The norm of $|\Psi_{\alpha,(t^0,p_t^0)}^{\rm{phy}}\rangle$ then becomes
\begin{align}
 ||\Psi_{\alpha,(t^0,p_t^0)}^{\rm{phy}} ||^2 
 &= \frac{1}{K} \int\limits_{\mathbb{R}} \mathrm{d}\lambda \, 
    \langle \hat{U}(\lambda) \Psi_{\alpha} |\Psi_{\alpha}\rangle 
 = \frac{2\pi\mu \hbar\omega_0 e^{-|\alpha|^2}}{K} 
 \sum\limits_{n=0}^{\infty}\frac{|\alpha|^{2n}}{n!}\epsilon_n^{\mu-1}|\Psi_{{t}^0,{p}_t^0}(-\epsilon^\mu_n)|^2 \nonumber \\
 &= \frac{2\pi\mu \hbar\omega_0 e^{-|\alpha|^2}}{K} 
 \sum\limits_{n=0}^{\infty}\frac{|\alpha|^{2n}}{n!}\epsilon_n^{\mu-1}|\Psi_{{t}^0-{p}_t^0}(\epsilon^\mu_n)|^2 \nonumber \\
& =\frac{2\pi e^{-|\alpha|^2}}{K} 
 \sum\limits_{n=0}^{\infty}c_{n;\mu}\frac{|\alpha|^{2n}}{n!}
\end{align}
here we used that for the absolute value we have $|\Psi_{{t}^0,{p}_t^0}(-\epsilon^\mu_n)|^2=|\Psi_{{t}^0,-{p}_t^0}(\epsilon^\mu_n)|^2$  and in the last line we defined $c_{n;\mu}:=\hbar\omega_0\mu\epsilon_n^{\mu-1}|\Psi_{{t}^0,-{p}_t^0}(\epsilon^\mu_n)|^2$. That the norm is finite is ensured by the fact that already $\frac{\alpha^n}{n!}$ is converging and  due to the absolute value of the Gaussian evaluated at $\epsilon_n^\mu$ for large values of $n$ the sum involved in the norm is even stronger decreasing. We can obtain normalised physical coherent states by choosing $K=2\pi$ and using the states
\begin{align}
\label{eq:normalisedPhysicalConstarinedCS}
|\widetilde{\Psi}_{\alpha,(t^0,p^0_t)}^{\rm{phy}}\rangle &:=\frac{|{\Psi}_{\alpha,(t^0,p^0_t)}^{\rm{phy}}\rangle}{||\Psi_{\alpha,(t^0,p_t^0)}^{\rm{phy}} ||}\nonumber \\
&=
\frac{e^{-\frac{|\alpha|^2}{2}}\int\limits_{\mathbb{R}}{\rm d}p_t \sum\limits_{n=0}^\infty \hbar\omega_0\mu\epsilon_n^{\mu-1}\delta(p_t+\epsilon_n^\mu)\frac{\alpha^n}{\sqrt{n!}}\Psi_{t^0,p_t^0}(p_t)|n\rangle \otimes |p_t\rangle }{\frac{2\pi e^{-|\alpha|^2}}{2\pi} 
 \sum\limits_{n=0}^{\infty}c_{n;\mu}\frac{|\alpha|^{2n}}{n!}}.
\end{align}
Then, we have
\begin{eqnarray*}
\langle \widetilde{\Psi}_{\alpha,(t^0,p^0_t)}^{\rm{phy}}\,|\, \widetilde{\Psi}_{\alpha,(t^0,p^0_t)}^{\rm{phy}}\rangle 
&=&e^{-|\alpha|^2}\sum\limits_{n=0}^\infty \hbar\omega_0\mu\epsilon_n^{\mu-1}\frac{|\alpha|^{2n}}{n!}\frac{\int\limits_{\mathbb{R}}
\mathrm{d}p_t\, \delta(p_t+\epsilon_n^\mu)|\Psi_{t^0,p_t^0}|^2(p_t)}{||\Psi_{\alpha,(t^0,p_t^0)}^{\rm{phy}} ||^2} \nonumber \\
&=&\frac{e^{-|\alpha|^2}\sum\limits_{n=0}^\infty c_{n;\mu}\frac{|\alpha|^{2n}}{n!}}{e^{-|\alpha|^2}\sum\limits_{n=0}^\infty c_{n;\mu}\frac{|\alpha|^{2n}}{n!}}=1.
\end{eqnarray*}
In case we compute the expectation value of the constraint $\hat{\widetilde{C}}$ with respect to the non-normalised physical coherent states $|\Psi_{\alpha,(t^0,p^0_t)}^{\rm{phy}}\rangle$ setting $K=2\pi$, we obtain
\begin{eqnarray*}
\langle\Psi_{\alpha,(t^0,p^0_t)}^{\rm{phy}}\,|\,\hat{\widetilde{C}}\,| \Psi_{\alpha,(t^0,p^0_t)}^{\rm{phy}}\rangle 
&=&e^{-|\alpha|^2}\sum\limits_{n=0}^\infty c_{n;\mu}\left(-\hbar\omega_0(n+\frac{1}{2})+(\epsilon_n^\mu)^{\frac{1}{\mu}}\right)\frac{|\alpha|^{2n}}{n!}=0.
\end{eqnarray*}
Next we compute the expectation value of the Dirac observable $\hat{H}_{\rm HO}$ in the physical coherent states and we end up with
\begin{align}
\frac{\langle {\Psi}_{\alpha,(t^0,p^0_t)}^{\rm{phy}}\,|\hat{H}_{\rm HO}\,| {\Psi}_{\alpha,(t^0,p^0_t)}^{\rm{phy}}\rangle}{||\Psi_{\alpha,(t^0,p_t^0)}^{\rm{phy}} ||^2}   
=\frac{\sum\limits_{n=0}^\infty c_{n;\mu}\hbar\omega_0(n+\frac{1}{2})\frac{|\alpha|^{2n}}{n!}}{\sum\limits_{n=0}^{\infty}c_{n;\mu}\frac{|\alpha|^{2n}}{n!}}
=\frac{\hbar\omega_0|\alpha|^2\sum\limits_{n=0}^\infty c_{n+1;\mu}\frac{|\alpha|^{2n}}{n!}}{\sum\limits_{n=0}^{\infty}c_{n;\mu}\frac{|\alpha|^{2n}}{n!}}+\frac{\hbar\omega_0}{2}.
\end{align}
As in \cite{Ashtekar:2005dm,Ashworth:1996yv} we assume that the coherent states are peaked on the constraint surface. If we further use that $|p^0_t|=E_0^{(s)}$, where $E_0^{(s)}$ denotes the energy of the original system we started with and the $0$-label was introduced here because the energy is determined by $q_0,p_0$, then we will have $\hbar\omega_0|\alpha|^2=|p^0_t|^{\frac{1}{\mu}}=(E_0^{(s)})^{\frac{1}{\mu}}$ yielding to 
\begin{equation}
\frac{\langle {\Psi}_{\alpha,(t^0,p^0_t)}^{\rm{phy}}\,|\hat{H}_{\rm HO}\,|{\Psi}_{\alpha,(t^0,p^0_t)}^{\rm{phy}}\rangle}{|| {\Psi}_{\alpha,(t^0,p^0_t)}^{\rm{phy}}||^2}     
=(E_0^{(s)})^{\frac{1}{\mu}}\frac{\sum\limits_{n=0}^\infty c_{n+1;\mu}\frac{|\alpha|^{2n}}{n!}}{\sum\limits_{n=0}^{\infty}c_{n;\mu}\frac{|\alpha|^{2n}}{n!}}+\frac{\hbar\omega_0}{2}.
\end{equation}
Although the zero point energy comes out exactly, in the case of the expectation value related to the classical energy $(E_0^{(s)})^\frac{1}{\mu}$ this is not like that. Here the corresponding contribution involves two sums one from the norm and a second one from the expectation value where the latter involves the coefficient $c_{n;\mu}$ with a shifted index by one. This carries over to a shift in $\epsilon_{n+1}^\mu=\lr\hbar\omega_0\lr n+\frac{3}{2}\rr\rr^{\mu}$ and to the absolute value of the Gaussian, thus we get
\begin{equation*}
c_{n+1;\mu}=\hbar\omega_0\mu\lr\hbar\omega_0\lr n+\frac{3}{2}\rr\rr^{\mu-1}|\psi_{t^0,-p_t^0}(\epsilon_{n+1}^\mu)|^2    
\end{equation*}
with 
\begin{equation*}
| \psi_{t^0,-p_t^0}(\epsilon_{n+1}^\mu)|^2=|C_{t^0,p^0_t,\hbar}|^2e^{-\frac{1}{(\hbar\sigma)^{2\mu}}(\epsilon_{n+1}^\mu + p_t^0))^2}.   
\end{equation*}
The sum  of the squared norm in the denominator involves the same expression but with $n$ and not $n+1$ in $c_{n+1;\mu}$. The label $p_t^0$ is the classical label of the coherent states associated with the temporal momentum. As discussed, on the classical constraint surface we can identify $\sgn(p^0_t)p^0_t$ with the classical energy being equal to $E_0^{(s)}$, that is the $\mu$-th  fractional power of the energy of the classical harmonic oscillator.  Due to the $\hbar$ in the denominator in the Gaussian it is narrowly peaked around the value of the classical energy $E_0^{(s)}$. Hence, the peak of the Gaussian with its fractional argument will be located at $(E_0^{(s)})$. Because the classical energy is assumed to be large compared to the eigenvalues $\hbar\omega_0(n+\frac{1}{2})$ a reasonable choice for $|p_t^0|=E_0^{(s)}$ is a value that corresponds to large $n$ in $\epsilon_n
^\mu$. Consequently, the peak and hence the main contribution of this Gaussian with fractional argument will be at large values for $n$. Furthermore, the $\frac{1}{n!}$ in each summand has the additional effect that the summands are further decreasing strongly with increasing $n$. Therefore, in the sum in the numerator we can replace $c_{n+1;\mu}$ by $c_{n;\mu}$ and the corrections due to this replacement are very tiny. The shift in $\epsilon_{n+1}^{\mu-1}$ involved in $c_{n+1;\mu}$ will be of minor order compared to the effects coming from the Gaussian and inverse factorial. If the absolute value of the Gaussian were absent, then to find a justification why large values of $n$ will be most dominant would be difficult. So we realise that this is a specific feature of the physical coherent states. Assuming that we choose reasonable values for the classical energy that are sufficiently large compared to the energy eigenvalues of the harmonic oscillator Hamiltonian we obtain
\begin{equation}
\label{eq:ExpE01}
\frac{\langle {\Psi}_{\alpha,(t^0,p^0_t)}^{\rm{phy}}\,|\hat{H}_{\rm HO}\,|{\Psi}_{\alpha,(t^0,p^0_t)}^{\rm{phy}}\rangle}{|| {\Psi}_{\alpha,(t^0,p^0_t)}^{\rm{phy}}||^2}     
=(E_0^{(s)})^{\frac{1}{\mu}}+\frac{\hbar\omega_0}{2}.
\end{equation}
Note that in \cite{Bolen:2004tq} a similar strategy was considered. There the physical inner product still involves integrals and therefore variables were introduced that encode the deviation from the value around which the Gaussians in the coherent states are peaked. The resulting semiclassical expectation values were then written as an expansion consisting of a classical momentum variable and the width of the Gaussian. Despite that this seems to be more elaborate than in our case in the sense that they also include corrections around the classical value, the techniques they use cannot directly be carried over to our case, since we have no integrals involving Gaussians for the expectation value with respect to physical states left. Furthermore, this corrections arise because functions in the integrand are Taylor expanded. More close to our case is the work done in  \cite{Ashtekar:2005dm} where among others semiclassical expectation values of generic observables being quadratic in annihilation and creation operators for constraints involving the number operator were discussed leading to a similar situation as in our case with two sums one from the norm in the denominator and the second one from the expectation values in the numerator. They considered the asymptotic values for these observables and assumed that one of the classical labels $\alpha_i$ is very large and tends to infinity. Given this, they could show that these two sums will drastically simplify, if they consider the dominant contributions yielding to the correct classical values of the quantum observables under these assumptions. 

Considering now the result in (\ref{eq:ExpE01}) we can solve this for the classical energy leading to an expression that involves fractional powers of the semiclassical expectation value of the harmonic oscillator Hamiltonian, namely 
\begin{equation}
\label{eq:ExpE02}
 E_0^{(s)} = \left(\langle {\Psi}_{\alpha,(t^0,p^0_t)}^{\rm{phy}}\,|\hat{H}_{\rm HO}\,| {\Psi}_{\alpha,(t^0,p^0_t)}^{\rm{phy}}\rangle -\frac{\hbar\omega_0}{2}\right)^\mu.   
\end{equation}
We realise that in the limit $\hbar\to 0$ the $\mu$-th power of the expectation value of $\hat{H}_{\rm HO}$ with respect to the normalised physical coherent states $|\widetilde{\Psi}_{\alpha,(t^0,p^0_t)}^{\rm{phy}}\rangle$ agrees with the classical energy $E_0^{(s)}$. If we had worked with the normal ordered Hamiltonian $:\hat{H}_{\rm HO}:$, as for instance done in \cite{Ashtekar:2005dm}, the $\hbar$ corrections due to the zero point energy would have even been absent. The reason why shifting the fractional power from the Hamiltonian to the temporal momentum works here. The fractional power gets reintroduced in the final result by requiring that for physical coherent states their labels are peaked on the constraint surface which is a physically reasonable assumption and in this sense carries the fractional power of the operators over to the classical labels of the coherent states where they can be handled in a simpler manner. Let us compare the situation at the physical and kinematical level in this aspect. For this purpose we consider the Dirac observable $|p_t|^{\frac{1}{\mu}}$ which at the physical level coincides with the harmonic oscillator Hamiltonian. In the two cases we obtain for the semiclassical expectation values
\begin{equation*}
 \frac{\langle {\Psi}_{\alpha,(t^0,p^0_t)}^{\rm{phy}}\,||\hat{p}_{t}|^{\frac{1}{\mu}}\,| {\Psi}_{\alpha,(t^0,p^0_t)}^{\rm{phy}}\rangle}{|| {\Psi}_{\alpha,(t^0,p^0_t)}^{\rm{phy}}||^2} =\frac{\hbar\omega_0|\alpha|^2\sum\limits_{n=0}^\infty c_{n+1;\mu}\frac{|\alpha|^{2n}}{n!}}{\sum\limits_{n=0}^{\infty}c_{n;\mu}\frac{|\alpha|^{2n}}{n!}}+\frac{\hbar\omega_0}{2}
 \approx \hbar\omega_0|\alpha|^2+\frac{\hbar\omega_0}{2}
 \end{equation*}
for large $n$ and
\begin{equation*}
\langle \widetilde{\Psi}_{\alpha,(t^0,p^0_t)}\,||\hat{p}_{t}|^{\frac{1}{\mu}}\,| \widetilde{\Psi}_{\alpha,(t^0,p^0_t)}\rangle
 =\frac{\Gamma(\frac{\frac{1}{\mu}+1}{2})}{\sqrt{\pi}}
\left((\hbar\sigma)^{2\mu}\right)^{\frac{1}{2\mu}}\kchf{-\frac{1}{2\mu},\frac{1}{2},-\frac{(p^0_t)^2}{(\hbar\sigma)^{2\mu}}}.
\end{equation*}
Thus, even if we assume that the coherent state is peaked on the classical constraint surface, where we can use that $|p_t^0|=E_0^{(s)}=(\hbar\omega_0|\alpha|^2)^\mu$ and consider the expansion of the Kummer function for large arguments shown in (\ref{eq:KummExp}), we observe that we obtain for the kinematical expectation values $\hbar$ corrections to $E_0^{(s)}$ that are not caused by the zero point energy of the harmonic oscillator but due to the in general fractional  power associated with the temporal momentum. The underlying reason for this is that in the case of the physical coherent states due to the involved delta function the inner product is modified and hence the in general fractional powers of $p_t$ need no longer to be integrated against the Gaussian of the coherent state which lead exactly to Kummer's function involved above in the kinematical case.

One can ask the question how the situation on the labels and the form of the states might change in case we apply reduced phase space quantisation instead of Dirac quantisation.
As pointed out in \cite{Ashtekar:2005dm} often the physical inner product can be identified with the inner product on the reduced phase space and we will discuss the situation for this model here. If we perform a reduced phase space quantisation, we can identify the phase space variable $t$ with our clock. Since the constraint $C=p_t+H^\mu_{\rm HO}$ is in deparametrised form, we can construct Dirac observables for $q,p$ by choosing a gauge fixing condition $G=t-\tau$ and use the power series expansion introduced in \cite{Vy1994,dittrich,dittrich2}. In this simple model the power series can be written in closed form and we obtain for the Dirac observables
\begin{align}
\label{eq:Obsqp}
O_q(\tau)&=\sum\limits_{n=0}^\infty \frac{(-1)^n(\tau-t)^n}{n!}\{q,H_{\rm HO}^\mu\}_{(n)},\quad 
O_p(\tau)=\sum\limits_{n=0}^\infty \frac{(-1)^n(\tau-t)^n}{n!}\{p,H_{\rm HO}^\mu\}_{(n)},  
\end{align}
where $\{f,g\}_{(n)}$ denotes the iterated Poisson bracket with $\{f,g\}_{(0)}=f$ and $\{f,g\}_{(n)}=\{\{f,g\}_{(n-1)},g\}$
with
\begin{align}
\{q,H_{\rm HO}^\mu\}_{(2n+1)}& = (-1)^n(\mu H_{\rm HO}^{\mu-1})^{2n+1}\omega_0^{2n+1}\frac{p}{m\omega_0},\quad
\{q,H_{\rm HO}^\mu\}_{(2n)} = (-1)^n(\mu H_{\rm HO}^{\mu-1})^{2n}\omega_0^{2n}q, \nonumber \\
\{p,H_{\rm HO}^\mu\}_{(2n+1)} &= (-1)^{n+1}(\mu H_{\rm HO}^{\mu-1})^{2n+1}\omega_0^{2n+1}m\omega_0 q,\quad
\{q,H_{\rm HO}^\mu\}_{(2n)} = (-1)^{n}(\mu H_{\rm HO}^{\mu-1})^{2n}\omega_0^{2n}p. 
\end{align}
Reinserting this back into the observables in (\ref{eq:Obsqp}) the closed form of these observables is given by
\begin{align}
 O_q(\tau)&=\sin(\mu H_{\rm HO}^{\mu-1}\omega_0(t-\tau))q+ \cos(\mu H_{\rm HO}^{\mu-1}\omega_0(t-\tau))\frac{p}{m\omega_0} ,\quad \nonumber \\
 O_p(\tau)&=\sin(\mu H_{\rm HO}^{\mu-1}\omega_0(t-\tau))p -\cos(\mu H_{\rm HO}^{\mu-1}\omega_0(t-\tau))m\omega_0 q.
\end{align}
The algebra of these observables satisfies the standard canonical Poisson algebra, that is $\{O_q,O_p\}=1$ and all remaining ones vanish. Given this explicit form that of the observables we can explicitly show that indeed the physical ${\rm H}_{\rm phys}=H^\mu_{\rm HO}(O_q,O_p)$ generates their evolution. We have 
\begin{align}
\frac{O_q(\tau)}{d\tau}&=-\mu H_{\rm HO}^{\mu-1}\omega_0\cos(\mu H_{\rm HO}^{\mu-1}\omega_0(t-\tau))q+ \mu H_{\rm HO}^{\mu-1}\omega_0\sin(\mu H_{\rm HO}^{\mu-1}\omega_0(t-\tau))\frac{p}{m\omega_0} ,\quad \nonumber \\
&= \mu H_{\rm HO}^{\mu-1}\frac{O_p(\tau)}{m}=\frac{\partial H^\mu_{\rm HO}(O_q,O_p)}{\partial O_p}
=\{O_q,H^\mu_{\rm HO}(O_q,O_p)\}
\end{align}
and
\begin{align}
 \frac{dO_p(\tau)}{d\tau}&=-\mu H_{\rm HO}^{\mu-1}\omega_0\cos(\mu H_{\rm HO}^{\mu-1}\omega_0(t-\tau))p -\mu H_{\rm HO}^{\mu-1}\omega_0\sin(\mu H_{\rm HO}^{\mu-1}\omega_0(t-\tau))m\omega_0 q.\nonumber \\
 &=-\mu H_{\rm HO}^{\mu-1}m\omega_0^2O_{q}=-\frac{\partial H^\mu_{\rm HO}(O_q,O_p)}{\partial O_q}
=\{O_p,H^\mu_{\rm HO}(O_q,O_p)\}
\end{align}
The physical Hamiltonian is ${\rm H}_{\rm phys}=H^\mu_{\rm HO}(O_q,O_p)$ can be quantised using the standard Schr\"odinger representation and hence the reduced phase space is just ${\cal H}^{\rm red}_{\rm phys}=L_2(\mathbb{R},{\rm d O}_q)$, where $\hat{O}_q$ acts by multiplication and $\hat{O}_p$ as a derivative operator. The quantisation of the Hamiltonian allows to formulate the corresponding Heisenberg equations for $\hat{O}_q$ and $\hat{O}_p$ with the Hamiltonian operator $\hat{H}^\mu_{\rm HO}$. Going over to the Schr\"odinger picture, one obtains a standard Schr\"odinger-like equations with $\hat{H}^\mu_{\rm HO}$ as the involved Hamiltonian operator. Physical coherent states on the reduced physical Hilbert space can be constructed as
\begin{align}
|\Psi_{O_\alpha}\rangle =e^{-\frac{|O_\alpha|^2}{2}}\sum\limits_{n=0}^\infty \frac{O_\alpha^n}{\sqrt{n!}}|n\rangle\quad{\rm with}\quad O_{\alpha}:=\sqrt{\frac{m\omega_0}{2\hbar}}O_{q_0}+i\sqrt{\frac{1}{2\hbar m\omega_0}}O_{p_0}.
\end{align}
With respect to the inner product of ${\cal H}^{\rm red}_{\rm phys}$ these physical coherent states are normalised as one can easily see. The physical coherent states obtained via group averaging can be isometrically embedded into ${\cal H}^{\rm red}_{\rm phys}$ using the map
\begin{align}
|n\rangle \to |\tilde{n}\rangle:=\sqrt{c_{n;\mu}}|n\rangle\quad {\rm with}\quad c_{n;\mu}:=\hbar\omega_0\mu\epsilon_n^{\mu-1}|\Psi_{{t}^0,-{p}_t^0}(\epsilon^\mu_n)|^2,\quad K:=2\pi,   
\end{align}
where we assumed, as mentioned above, that the constant $C_{t^0,p^0_t,\hbar}$ was chosen such that the coherent states $\Psi_{t^0,p_{t^0}}$ in $L_2(\mathbb{R},{\rm dp}_t)$ were normalised. Using these rescaled states $|\tilde{n}\rangle$ in the reduced inner product  yields the same result like for the physical coherent states in the physical inner product.  Because in the reduced phase space any function involving the variables $(t,p_t)$ can be expressed as a function of $O_q,O_p$ only the expectation values for Dirac observables with respect to physical coherent states using group averaging and reduced phase space quantisation agree under the identification $q_0\to O_{q_0}, p_0 \to O_{p_0}$. 

Finally, let us briefly summarise the results obtained in this section. We used an Euler rescaling to rewrite the deparametrised constraint in the form that it involves the Hamiltonian linearly and the temporal momentum with some in general fractional power. We showed that using Kummer's confluent hypergeometric function the standard coherent states yield a good semiclassical approximation of the quantum constraint operator at the kinematical level. Then we applied group averaging to construct physical coherent states that are assumed to be peaked on the classical constraint surface. The latter allows to relate the absolute values of the labels of the coherent states to the energy of the system represented by the temporal momentum at the classical level. In this sense the coherent states intrinsically encode some dynamical properties via their labels and are beside the group averaging adapted to the constraint under consideration. Note that using coherent states that are peaked on the constraint surface was also crucial in \cite{Ashtekar:2005dm} in order to obtain good semiclassical results for the operators corresponding to the classical Dirac observables.  In our example discussed so far the coherent states are perfectly adapted to the Hamiltonian $\hat{H}_{\rm HO}$. As a consequence the relation between the classical energy $E_0^{(s)}$ and the semiclassical expectation value in (\ref{eq:ExpE01}) and (\ref{eq:ExpE02}) is very simple. For more complicated Hamiltonians one obtains a more complicated function of the coherent states labels $\alpha$ in which one then also replaces $\hbar\omega_0|\alpha|^2$ by $(E_0^{(s)})^{\frac{1}{\mu}}$. However, in order for the semiclassical states to be reasonable in lowest order in $\hbar$, we expect to obtain $(E_0^{(s)})^{\frac{1}{\mu}}$ plus possible further additional terms which then come with higher orders in $\hbar$ and can be interpreted as small corrections to the classical value. In order to test whether the semiclassical limit is correct, which corresponds to the limit $\hbar\to0$, this method here can be useful but to work with the possible corrections involved could become problematic because the final step involves solving for $E_0^{(s)}$ which requires that the inverse function of the right hand side of (\ref{eq:ExpE02}) exists. 

If we want to encode that the coherent states are peaked on the constraint surface directly into their labels, we can achieve this by implementing the corresponding restriction on the labels $\alpha$. In our case we have $\hbar\omega_0|\alpha|^2=(E_0^{(s)})^{\frac{1}{\mu}}$. Hence, we can label the coherent states with $\alpha=\frac{|E_0^{(s)}|^{\frac{1}{2\mu}}}{\sqrt{\hbar\omega_0}}e^{i\varphi}$. Then following the computations done above, we also end up with the results in (\ref{eq:ExpE01}) and (\ref{eq:ExpE02}). Although the states are adapted to the fractional power $\mu$ of the Hamiltonian by construction the label involves the inverse power $\frac{1}{\mu}$ which requires to solve for $E_0^{(s)}$. In the next section we want to consider the aspect that the labels of the coherent states carry some dynamical information from a different perspective and show that one can use coherent states for which the semiclassical expectation values involve directly the $\mu$th power of the classical energy and not the inverse power $\frac{1}{\mu}$.

\section{Coherent states in constrained systems: Reduced quantisation and generalised coherent states for fractional Hamiltonians}
\label{sec:GenCS}
In the last section we discussed how to apply the formalism developed in \cite{Ashworth:1996yv,Ashtekar:2005dm} and combine it with the Euler rescaling in the context of an extended phase space to obtain coherent states which are, in the sense discussed above,  adapted to square root Hamiltonians  or more general fractional Hamiltonians. In this section we want to address the question of appropriate coherent states for fractional Hamiltonians from a different angle. As we saw in the last section the physical coherent states differ from the kinematical ones by a restriction on their label set that is determined by the form of the constraint under consideration. Following this route here, we want to incorporate already into the construction of the coherent states that they should be well suited for fractional powers of the Hamiltonian. For this purpose we can restrict our discussion to the case of reduced phase space quantisation and hence do not consider the degrees of freedom corresponding to $t,p_t$ in the extended phase space here, since we have already shown in the last section that we obtain similar results for Dirac and reduced quantisation for the example that we consider in this article. In this case we quantise the algebra of Dirac observables shown in (\ref{eq:Obsqp}) in the standard Schr\"odinger representation and their dynamics in the Heisenberg picture is generated by $\hat{H}_{\rm HO}^\mu$, the operator corresponding to the physical Hamiltonian of the Dirac observables.

There exist already preliminary work in the literature in the framework of so-called fractional Poisson distributions \cite{2003:Laskin1,2009:Laskin2}, where in \cite{2009:Laskin2} generalised coherent states were constructed based on functions denoted as Mittag-Leffler functions which will be defined below in (\ref{eqDefMittagLefflerFunction}). The work in \cite{2009:Laskin2} analyses in detail the properties of these coherent states and presents a proof for their resolution of identity and we will briefly review the introduction of these states in section \ref{sec:ReviewLaskinStates}. As we will show in subsection \ref{sectionGeneralisedCoherentStatesForFractionalHamiltonians} the proof presented in \cite{2009:Laskin2} is based on an incorrect assumption as far as the orthogonality of the angular part of the coherent states is considered. By generalising the measure involved in the resolution of identity along the lines introduced in \cite{Klauder:1995yr} we can correct this and introduce a slightly different set of coherent states that satisfies a resolution of identity. Furthermore, the set of coherent states introduced here, has the property that the states are still eigenstates of the annihilation operator which is is not the case for the coherent states in \cite{2009:Laskin2}. In subsection \ref{sec:PolyHam} we will construct a similar type of coherent states in the framework of polymerised Hamiltonians which might be the more suitable framework for these kind of states. Finally in subsection \ref{sectionSemiclassPerturbationFractionalHamiltonians} we compare our results with those obtained from the AQG algorithm for this model under consideration.

\subsection{Coherent States based on the fractional Poisson distribution}
\label{sec:ReviewLaskinStates}
Before we introduce the generalised set of coherent states we briefly review the main results from \cite{2003:Laskin1,2009:Laskin2} because part of them can be seen as the motivation for introducing the generalised harmonic oscillator coherent states in this work. One of the main ideas in this construction is to obtain states that are no longer build from a Poisson distribution, like the standard harmonic oscillator coherent states, but a more general probability distribution associated with the Mittag-Leffler function. This function can be understood as a generalisation of the exponential function usually involved in the Poisson distribution. There exist several generalisations of the original Mittag-Leffler function which are encoded in additional parameters the function depends on. The original Mittag-Leffler function just depends on one parameter $\mu>0$ and is given by
\begin{equation}
\label{eqDefMittagLefflerFunction}
z\in\mathbb{C},\quad z\mapsto E_{\mu}(z):=\sum\limits_{k=0}^\infty \frac{z^k}{\Gamma(\mu n+1)},   
\end{equation}
where $\Gamma$ denotes the standard Gamma function with $\Gamma(z+1)=z\Gamma(z)$ and  $E_\mu$ is an entire function. It can be understood as a kind of stretched exponential due to the Gamma function in the denominator. In the special case of $\mu=1$ we have $\Gamma(n+1)=n!$ and then the Mittag-Leffler function $E_\mu$ becomes the usual exponential function. In this work we are interested in the parameter range $0<\mu\leq 1$. The coherent states introduced in \cite{2009:Laskin2} are of the form
\begin{equation}
\label{eq:LaskinCS}
|\varsigma ;\mu\rangle_{\rm ML} = \sum\limits_{n=0}^\infty \frac{(\sqrt{\mu}\varsigma^\mu)^n}{\sqrt{n!}}\left(E^{(n)}_{\mu}(-\mu |\varsigma|^{2\mu})\right)^{\frac{1}{2}}|n\rangle, \quad 0<\mu\leq 1,
\end{equation}
where we introduced the label $\rm ML$ to emphasise that the states involve the Mittag-Leffler function and we introduced $\varsigma=\sqrt{\frac{m\omega_0}{2\hbar}}q_0+i\sqrt{\frac{1}{2\hbar m\omega_0}}p_0$, where $E^{(n)}_{\mu}(-\mu |\varsigma|^{2\mu})$ denotes the n-th derivative of $E_\mu$ given by
\begin{equation*}
 E^{(n)}_\mu(-\mu|\varsigma|^{2\mu}):=\frac{d^n}{dz^n}E_{\mu}(z)\Big|_{z=-\mu|\varsigma|^{2\mu}}.   
\end{equation*}
For the choice of $\mu=1$ they reduce to the standard harmonic oscillator coherent states with the identification $\varsigma=\alpha$
\begin{align*}
|\alpha ;1 \rangle_{\rm ML} &= \sum\limits_{n=0}^\infty \frac{\alpha^n}{\sqrt{n!}}\left(E^{(n)}_{1}(-|\alpha|^2)\right)^{\frac{1}{2}}|n\rangle
=\sum\limits_{n=0}^\infty \frac{\alpha^n}{\sqrt{n!}}\left(e^{-|\alpha|^2}\right)^{\frac{1}{2}}|n\rangle \\
&=e^{-\frac{1}{2}|\alpha|^2}\sum\limits_{n=0}^\infty \frac{\alpha^n}{\sqrt{n!}}|n\rangle.
\end{align*}
As shown in \cite{2009:Laskin2} the generalised coherent states in (\ref{eq:LaskinCS}) are normalised. Furthermore, in \cite{2009:Laskin2} it is claimed that these coherent states satisfy a resolution of identity. However, the proof presented in \cite{2009:Laskin2} involves a mistake and we will discuss below how such mistake can be avoided by modifying the measure involved in the resolution of identity. This will then provide the basis for introducing a generalisation of the harmonic oscillator coherent states that are better adapted to operators involving fractional powers.

Now the reason why nevertheless these states are interesting in the context of fractional Hamiltonians is that in these cases the expectation value of the number operator $\hat{n}=\hat{a}^\dagger\hat{a}$ is a fractional power of $|\varsigma|^2=|\alpha|^2$, where the last equality is obtained by comparison of
\begin{equation*}
{}_{\rm ML}\langle \varsigma;\mu|\,\had \ha\, |\varsigma;\mu\rangle_{\rm ML} =  \frac{\mu|\varsigma|^{2\mu}}{\Gamma(\mu+1)},    
\end{equation*}
with the expectation value $\langle\alpha|\,\hat{a}^\dagger\hat{a}\, |\alpha\rangle=|\alpha|^2$ for the standard coherent states which will again be recovered if we set $\mu=1$ in the general case. These properties look already interesting as far as fractional operators are considered, however due to the factor coming from the Gamma function, the semiclassical limit might be stretched as well and hence deviates from the correct expression by this factor. Moreover, a further differences of $|\varsigma;\mu\rangle_{\rm ML}$ compared to the standard coherent states $|\alpha\rangle$ is that the generalised states for $\mu\not=1$ are no longer eigenstates of the annihilation operator $\hat{a}$. The reason for this is that the n-th derivative of the Mittag-Leffler function $E^{(n)}_\mu$ depends on the order of $n$ and thus cannot just be pulled in front of the summation, as it is the case for the standard exponential, whose derivative for all orders of $n$ involves again the exponential function only up to possible additional factors coming from inner derivative contributions. Furthermore, since $|\varsigma; \mu\rangle_{\rm ML}$ are no eigenstates of $\hat{a}$, these states are less suitable for other operators than the number operator which have more generic dependencies on $\hat{a}$ and $\hat{a}^\dagger$ such as for instance a polynomial one. Note that there exists a generalised annihilation operator of the form
\begin{equation*}
\hat{a}_{(\mu)}|n\rangle =\sqrt{g(n;\mu)}|n-1\rangle\quad{\rm with}\quad g(n;\mu)=\sqrt{\frac{n E^{(n-1)}_\mu(-\mu|\varsigma|^{2\mu})}{E^{(n)}_\mu(-\mu|\varsigma|^{2\mu})}}   
\end{equation*}
for which $|\varsigma;\mu\rangle_{\rm ML}$ is an eigenstate with eigenvalue $\sqrt{\mu}\varsigma^\mu$. For the choice $\mu=1$ the operator $\hat{a}_{(\mu)}$ becomes the standard annihilation operator because $g(n;1)=\sqrt{n}$. Also only for this choice the algebra of  $\hat{a}_{(\mu)}$, $\hat{a}^\dagger_{(\mu)}$ and the identity operator satisfy the standard commutation relations, in general it is more complicated and given by
\begin{equation}
[\hat{a}_{(\mu)},\hat{a}_{(\mu)}^{\dagger}]|n\rangle 
=\left(\frac{(n+1)E^{(n)}_\mu(-\mu|\varsigma|^{2\mu})}{E^{(n+1)}_\mu(-\mu|\varsigma|^{2\mu})}
-
\frac{n\,E^{(n-1)}_\mu(-\mu|\varsigma|^{2\mu})}{E^{(n)}_\mu(-\mu|\varsigma|^{2\mu})}\right)|n\rangle, 
\end{equation}
and even depends on the state $|n\rangle$. Here we used that $\hat{a}^\dagger_{(\mu)}|n\rangle=\sqrt{g(n+1;\mu)}|n+1\rangle$.

Given this, in the next section we want to discuss a different set of generalised coherent states which are also normalised, satisfy a resolution of identity but in addition are also eigenstates of the annihilation operator $\hat{a}$ with in general eigenvalues of fractional powers of $\alpha$. These states are different from the ones described in \cite{2009:Laskin2}, since they do not involve the general Mittag-Leffler function $E_\mu$  but the Mittag-Leffler function $E_1$ only which agrees with the exponential function. 
They can be understood as standard coherent states of the harmonic oscillator but with labels that have been adopted to the fractional Hamiltonian under consideration. 
The reason why we want to construct these states in the case of fractional powers of the Hamiltonian is that given these states we can consider the standard harmonic oscillator Hamiltonian as a kind of effective Hamiltonian for the computation of the semiclassical expectation values. This is the case because the coherent states are constructed in a way that they encode the properties of the  fractional operator. 
\subsection{Generalised Coherent States for fractional Schrödinger Hamiltonians}
\label{sectionGeneralisedCoherentStatesForFractionalHamiltonians}
The generalised coherent states that will be discussed in this section are given by
\begin{equation}
\label{eq:GenCS}
|\alpha;\mu\rangle =\sum\limits_{n=0}^\infty \frac{(\alpha^\mu)^n}{\sqrt{n!}}e^{-\frac{1}{2}|\alpha|^{2\mu}}|n\rangle 
=\sum\limits_{n=0}^\infty \frac{(\alpha^\mu)^n}{\sqrt{n!}}\left(E^{(n)}_1(-|\alpha|^{2\mu})\right)^{\frac{1}{2}}|n\rangle
\end{equation}
for $0<\mu\leq 1$ and we used that $E_1(z)=e^{z}$. We trivially rewrote $|\alpha;\mu\rangle$ in the last step only to make the relation to the states $|\varsigma;\mu\rangle_{\rm ML}$ in (\ref{eq:LaskinCS}) more transparent. 
Likewise to the states in (\ref{eq:LaskinCS}) these states depend on an additional parameter $\mu$ but their explicit dependence is different. Moreover, we keep the exponential function in the definition and do not consider the Mittag-Leffler function here. The latter ensures that these states are still eigenstates of the usual annihilation operator. If we compare the corresponding probability distributions corresponding to the states $|\varsigma;\mu\rangle_{\rm ML}$ from section \ref{sec:ReviewLaskinStates}   and $|\alpha;\mu\rangle$ we obtain
\begin{equation}
\label{eq:PmuLaskin}
P^{\varsigma,{\rm ML}}_\mu(n):=|\langle n |\varsigma;\mu\rangle_{\rm ML}|^2=\frac{(\mu|\varsigma|^{2\mu})^n}{n!}\frac{d^n}{dz^n}E_{\mu}(z)\Big|_{z=-\mu|\varsigma|^{2\mu}}   
\end{equation}
and 
\begin{equation}
\label{eq:Pmu}
P^\alpha_\mu(n):=|\langle n |\alpha;\mu\rangle|^2=\frac{(|\alpha|^{2\mu})^n}{n!}e^{-|\alpha|^{2\mu}}=\frac{(|\alpha|^{2\mu})^n}{n!}\frac{d^n}{dz^n}E_1(z)\Big|_{z=-|\alpha|^{2\mu}}\, ,  
\end{equation}
where we as above in (\ref{eq:GenCS}) rewrote $P^\alpha_\mu(n)$ in the last step only to show their exact relation to $P^\varsigma_\mu(n)$. As shown in \cite{2009:Laskin2} the probability distribution $P^\varsigma_\mu(n)$ has the mean value 
\begin{equation*}
\overline{n}_{\varsigma,{\rm ML}}=\sum\limits_{n=0}^\infty n P^{\varsigma,{\rm ML}}_\mu(n)= \frac{\mu|\varsigma|^{2\mu}}{\Gamma(\mu+1)}. 
\end{equation*}
Considering the action of the annihilation and creation operator as 
\begin{equation*}
\hat{a}|n\rangle=\sqrt{n}|n-1\rangle,\quad \hat{a}^\dagger|n\rangle=\sqrt{n+1}|n+1\rangle
\end{equation*}
one can show that this is equal to  \cite{2009:Laskin2}
\begin{equation}
\label{eq:nMeanPLaskin}
\overline{n}_\varsigma=\sum\limits_{n=0}^\infty n P^\varsigma_\mu(n)=  \frac{\mu|\varsigma|^{2\mu}}{\Gamma(\mu+1)}=\langle\varsigma;\mu |\,\hat{a}^\dagger\hat{a}\,| \varsigma;\mu\rangle   
\end{equation}
which is the relevant form for our physical applications. If we perform the same computations for $P^\alpha_\mu(n)$ and the states $|\alpha;\mu\rangle$, we will end up with
\begin{equation}
\label{eq:nMean}
\overline{n}_\alpha=\sum\limits_{n=0}^\infty n P^\alpha_\mu(n)=|\alpha|^{2\mu}=\langle\alpha;\mu |\,\hat{a}^\dagger\hat{a}\,| \alpha;\mu\rangle.  
\end{equation}
Despite that the final results in (\ref{eq:nMeanPLaskin}) and (\ref{eq:nMean}) look similar, the way one obtains them is different. In the first case the states $|\varsigma;\mu\rangle_{\rm ML}$ are no eigenstates of $\hat{a}$ but if one computes the summation in (\ref{eq:nMeanPLaskin}) one has to combine the sum over $n$ with the sum over $k$ involved in the derivatives of the Mittag-Leffler function and uses the binomial theorem. The latter absorbs one of the sums and the second runs over the power index of the binomial theorem. However, the arguments inside the bracket in the binomial theorem are just identical up to a sign so that the only non-vanishing contributions comes from the case where the power index is equal to zero, see \cite{2003:Laskin1,2009:Laskin2} for more details. The combination of the two sums is only possible at the level of expectation values because when we consider the action on an individual coherent state the sum over $k$ is still inside a square root and thus cannot be combined with the outer sum over $n$. 

On the other hand for $P^\alpha_\mu(n)$ we can use that $|\alpha;\mu\rangle$ is an eigenstate of $\hat{a}$, which can be easily shown by
\begin{align}
\label{eq:Proofeigenstate}
\hat{a}|\alpha;\mu\rangle &=e^{-\frac{1}{2}|\alpha|^{2\mu}}\sum\limits_{n=1}^\infty\frac{(\alpha^\mu)^n}{\sqrt{n!}}\sqrt{n}|n-1\rangle
=e^{-\frac{1}{2}|\alpha|^{2\mu}}\sum\limits_{n=0}^\infty\frac{(\alpha^\mu)^{n+1}}{\sqrt{n!}}|n\rangle \\ \nonumber
&=\alpha^\mu |\alpha;\mu\rangle.
\end{align}
Hence the eigenvalue is just given by $\alpha^\mu$. 
Let us check that the states $|\alpha;\mu\rangle$ satisfy all three requirements: (i) normalisation, (ii) satisfy a resolution of identity and (iii) are eigenstates of the annihilation operator $\hat{a}$, where the last one was already shown above. 
The normalisation can easily  be shown by
\begin{equation*}
(i)\, \langle \alpha;\mu \, |\, \alpha ;\mu\rangle  =
\sum\limits_{n=0}^\infty P^\alpha_\mu(n)=e^{-|\alpha|^{2\mu}}\sum\limits_{n=0}^\infty \frac{(|\alpha|^{2\mu})^n}{n!}
=e^{-|\alpha|^{2\mu}}e^{|\alpha|^{2\mu}}=1.
\end{equation*}
In order that these states qualify as coherent states the continuity in the parameter $\alpha$ needs to be given, see \cite{Klauder+Skagerstam:1985}. This is trivially satisfied here because $\alpha$ is the usual coherent states label used for the harmonic oscillator coherent states. The usual overcompleteness relation for the harmonic oscillator coherent states generalises to 
\begin{align*}
\langle \alpha ; \mu\, |\, \beta; \mu\rangle=\exp(-\frac{1}{2}\left(|\alpha|^{2\mu}+|\beta|^{2\mu}-2(\alpha^*\beta)^\mu\right)),     
\end{align*}
that will yield the usual expression if we set $\mu=1$. 

As far as  (ii) the resolution of identity is considered for the conventional harmonic oscillator coherent states we have 
\begin{eqnarray*}
\int\limits_{\mathbb{C}} \mathrm{d}^2\alpha\, |\alpha\rangle \langle \alpha | W_\mu(|\alpha|^2)=\uop,\quad {\rm with}\quad W_\mu(|\alpha|^2)=\frac{1}{\pi}
\end{eqnarray*}
where $\mathrm{d}^2\alpha=\mathrm{d}(\Re{(\alpha)}\mathrm{d}(\Im{(\alpha)})$. This can be proven by transforming   $\alpha=\sqrt{\frac{m\omega_0}{2\hbar}}q_0+i\sqrt{\frac{1}{2\hbar m\omega_0}}p_0$ to polar coordinates yielding $\alpha=\rho e^{i\phi}$ with $\rho:=|\alpha|$  with $0\leq \rho < \infty$, $0\leq \phi < 2\pi$ and using that $\{e^{in\phi}\}_{n\in\mathbb{N}}$ is an orthonormal basis in $L_2([0,2\pi], \frac{d\phi}{2\pi})$. For the coherent states $|\alpha ; \mu\rangle _{\rm ML}$ in \cite{2009:Laskin2} as well as the ones $|\alpha ; \mu\rangle$ introduced in our work, we work with a fractional label $\alpha^\mu$ and hence we have $\alpha^\mu=\rho^\mu e^{in\mu\phi}$. Thus, for $\mu\not=1$ $\{e^{in\mu\phi}\}_{n\in\mathbb{N}}$ is no longer an orthonormal basis of $L_2([0,2\pi], \frac{d\phi}{2\pi})$. However, this seems to be have overseen in the proof presented in \cite{2009:Laskin2} which therefore cannot be performed in the way presented in \cite{2009:Laskin2} and yields to the incorrect conclusion that these states satisfy a resolution of identity. As we will show this issue can be circumvented by generalising the measure that is involved in the resolution of identity along the lines introduced in \cite{Klauder:1995yr} and for instance applied in \cite{Fox:1999} and will use this strategy to prove that the states $|\alpha ; \mu\rangle$ satisfy a resolution of identity. For this purpose, as suggested in \cite{Klauder:1995yr}, we extend the polar coordinates to their covering space with the domains $0\leq \rho <\infty$ and $-\infty <\phi < \infty$ and consider a measure $\nu(\rho,\phi)$ defined by
\begin{equation*}
\int\limits \mathrm{d}\nu(\rho,\varphi;\mu) F(\rho,\phi) := \lim\limits_{\Gamma\to \infty}\frac{1}{2\Gamma} \int\limits_{0}^{\infty}\,\mathrm{d}\rho\, W_{\mu}(\rho^2)\int\limits_{-\Gamma}^{\Gamma}\mathrm{d}\,\phi F(\rho,\phi), 
\end{equation*}
where $W_\mu(\rho^2)$ is a still to be determined positive weight function.  This yields
\begin{align*}
&\int\limits \mathrm{d}\nu(|\alpha|,\phi;\mu) \, |\alpha;\mu\rangle \langle \alpha ;\mu|\\
&=\lim\limits_{\Gamma\to\infty}\frac{1}{2\Gamma}\int\limits_0^\infty \, \mathrm{d}|\alpha|\, 
W_\mu(|\alpha|^2)\int\limits_{-\Gamma}^{\Gamma} \mathrm{d}\phi  |\alpha|,\phi;\mu\rangle \langle |\alpha|,\phi ;\mu|
\\
&=  
\lim\limits_{\Gamma\to \infty}\frac{1}{2\Gamma}\int\limits_{-\Gamma}^{\Gamma} \mathrm{d}\phi\, \int\limits_0^\infty \mathrm{d}\rho\, \sum\limits_{n,m=0}^\infty e^{i\mu(n-m)\phi} \frac{\rho^{\mu(n+m)+1}}{\sqrt{n!}\sqrt{m!}}e^{-\rho^{2\mu}} W_\mu(\rho^2)|n\rangle\langle m| \\
&=
\lim\limits_{\Gamma\to\infty}\frac{1}{2\Gamma}\sum\limits_{n,m=0}^\infty\int\limits_0^\infty \mathrm{d}\rho\, \frac{\rho^{\mu(n+m)+1}}{\sqrt{n!}\sqrt{m!}}e^{-\rho^{2\mu}} W_\mu(\rho^2) \int\limits_{-\Gamma}^{\Gamma} \mathrm{d}{\phi}\,  e^{i(n-m)\mu{\phi}}|n\rangle\langle m|, 
\end{align*}
where in the last step we interchanged the order of summation and integration. Now we can use that $\{e^{is\phi}\}_{s\in\mathbb{R}}$ is an orthonormal basis in $L_2(\mathbb{R}_{\rm Bohr},\mu_{\rm Bohr})$ where the inner product of this Hilbert space  can be expressed as $\langle f\, ,\, g\rangle=\lim\limits_{\Gamma\to\infty}\frac{1}{2\Gamma}\int_{-\Gamma}^{\Gamma} \,\mathrm{d}\phi\overline{f}(\phi)g(\phi)$. Performing the integration over the angle ${\phi}$ we obtain
\begin{align*}
&\int\limits \mathrm{d}\nu(|\alpha|,\phi;\mu) \, |\alpha;\mu\rangle \langle \alpha ;\mu|
= 
\sum\limits_{n=0}^\infty\int\limits_0^\infty \mathrm{d}\rho\, \frac{\rho^{2\mu n+1}}{n!}e^{-\rho^{2\mu}}W_\mu(\rho^2)|n\rangle\langle n| \\
&=
\frac{1}{2}\sum\limits_{n=0}^\infty\int\limits_0^\infty \mathrm{d}x\, \frac{x^{\mu n}}{n!}e^{-x^{\mu}}W_\mu(x)|n\rangle\langle n|,
\end{align*}
where in the last step we used the variable substitution $x=\rho^2$. Now we apply a further change of variables and introduce $y=x^\mu$ with $dy=\mu x^{\mu-1}dx=\mu y y^{-\frac{1}{\mu}}dx$. This results in
\begin{align*}
\int\limits \mathrm{d}^2\alpha\, |\alpha;\mu\rangle \langle \alpha ;\mu| W_\mu(|\alpha|^2)
= \frac{1}{2}\sum\limits_{n=0}^\infty\frac{1}{n!}\int\limits_0^\infty \mathrm{d}y\, y^n e^{-y} \frac{\mu y^{\frac{1}{\mu}}}{y}W_\mu(y)|n\rangle\langle n|. 
\end{align*}
Now we choose the weight function to be
\begin{align*}
W_\mu(y)=\frac{2y}{\mu y^{\frac{1}{\mu}}} \quad \longrightarrow W_\mu(\rho^2)=\frac{2}{\mu}(\rho^{2})^{\frac{\mu-1}{\mu}},
\end{align*}
which is positive, i.e.\,$W_\mu(\rho^2)>0$, then we end up with
\begin{equation}
\begin{split}
 (ii)\, &\int\limits \mathrm{d}\nu(|\alpha|,\phi;\mu) \, |\alpha;\mu\rangle \langle \alpha ;\mu|
 =
 \sum\limits_{n=0}^\infty\frac{1}{n!}\int\limits_0^\infty \mathrm{d}y\, y^ne^{-y}|n\rangle\langle n| \nonumber \\
 & 
 =
 \sum\limits_{n=0}^\infty\frac{\Gamma(n+1)}{n!}|n\rangle\langle n|
 = \sum\limits_{n=0}^\infty|n\rangle\langle n|=\uop
 \end{split}
 \end{equation}
and this proves the resolution of identity for the states $|\alpha ; \mu\rangle$. For the special choice of $\mu=1$ they weight function reduces to $W_\mu(\rho^2)=2$ which is exactly the weight function one obtains for the standard harmonic oscillator coherent states in case one performs a similar generalisation of the measure for the angular part as we did above.

Given the states $|\alpha;\mu \,\rangle$ labelled by $\mu$, let us discuss how we can use them as semiclassical states for operators involving fractional powers. We turn back to our example where the physical Hamiltonian on the reduced phase is given by $H_{\rm HO}^\mu$. Now in the quantum theory we consider as an integer power substitute for the Hamiltonian operator $\hat{H}_{\rm HO}^\mu$, the operator
\begin{align}
\label{eq:SubHO}
\widetilde{\hat{H}}_{\rm HO}:=(\hbar\omega_0)^{\mu-1}\hat{H}_{\rm HO}=(\hbar\omega_0)^\mu \left(\hat{a}^\dagger\hat{a}+\frac{\uop}{2}\right). 
\end{align}
Considering the generalised coherent states $|\alpha;\mu\rangle$ above for the semiclassical expectation value we obtain
\begin{align}
    \label{eq:ExpGenCSH}
\langle\alpha;\mu\, |    \widetilde{\hat{H}}_{\rm HO}\, | \alpha ;\mu\rangle
&=\langle \alpha ;\mu\, |  (\hbar\omega_0)^{\mu}\left(\hat{a}^\dagger\hat{a}+\frac{\uop}{2}\right)\, | \alpha ;\mu\rangle \nonumber\\
&=(\hbar\omega_0)^{\mu}|\alpha|^{2\mu}+\frac{(\hbar\omega_0)^\mu}{2}\nonumber \\
&=E_0^{\mu}+\frac{(\hbar\omega_0)^\mu}{2}.
\end{align}
From the last line we immediately see that up to the zero point energy that vanishes in the $\hbar\to 0$ limit the expectation value of the substitute operator, which only involves integer powers of $\hat{H}_{\rm HO}$, yields the correct classical limit in the zeroth order of $\hbar$. Following this route for different fractional powers of the harmonic oscillator Hamiltonian, we can always use $\hat{H}_{\rm HO}$ as a substitute operator for $\hat{H}_{\rm HO}^{\mu}$ supposed that we multiply $\hat{H}_{\rm HO}$ with the appropriate fractional powers of $\hbar\omega_0$ for dimensional reasons. For the fluctuations we obtain with
\begin{align}
\langle\alpha;\mu\, | (\widetilde{\hat{H}}_{\rm HO})^2\, |\alpha;\mu\rangle 
&=
\langle\alpha;\mu\, | (\hbar\omega_0)^{2\mu}\hat{H}_{\rm HO}^2\, |\alpha;\mu\rangle 
=(\hbar\omega_0)^{2\mu}(|\alpha|^{4\mu}+2|\alpha|^{2\mu}+\frac{1}{4})
\end{align}
the expected result
\begin{align}
(\Delta\widetilde{\hat{H}}_{\rm HO})^2=(\hbar\omega_0)^{2\mu}|\alpha|^{2\mu}=(\hbar\omega_0)^\mu E_0^\mu.    
\end{align}
These fluctuations come with a non-vanishing fractional power of $\hbar$ and are thus small compared to $E_0$ and vanish in the $\hbar\to 0$ limit.

If we use the coherent states introduced by Laskin for the same expectation values, as shown in \cite{2003:Laskin1}, we will end up with  
\begin{align}
{}_{\rm ML}\langle\varsigma;\mu\, | \widetilde{\hat{H}}_{\rm HO}\, |\varsigma;\mu\rangle_{\rm ML} &=  (\hbar\omega_0)^\mu(\frac{\mu|\varsigma|^{2\mu}}{\Gamma(\mu+1)}+\frac{1}{2})=\frac{\mu E_0^\mu}{\Gamma(\mu+1)}+\frac{(\hbar\omega_0)^\mu}{2}    
\end{align}
showing that even in the lowest order of $\hbar$ we do not obtain the expected classical limit if $\mu\not=1$. For the fluctuations following \cite{2003:Laskin1} we use that 
\begin{align}
{}_{\rm ML}\langle\varsigma;\mu\, | (\widetilde{\hat{H}}_{\rm HO})^2 \, |\varsigma;\mu\rangle_{\rm ML} = 
2(\hbar\omega_0)^{\mu}\frac{\mu E_0^\mu}{\Gamma(\mu+1)}+\left(\frac{\mu E_0^\mu}{\Gamma(\mu+1)}\right)^2\left(\frac{\sqrt{\pi}\Gamma(\mu+1)}{2^{2\mu-1}\Gamma(\mu+\frac{1}{2})}\right)+\frac{(\hbar\omega_0)^{2\mu}}{4}
\end{align}
and this leads to
\begin{align}
 (\Delta\widetilde{\hat{H}}_{\rm HO})^2_{\varsigma,{\rm ML}}=(\hbar\omega_0)^{\mu}\frac{\mu E_0^\mu}{\Gamma(\mu+1)}+\left(\frac{\mu E_0^\mu}{\Gamma(\mu+1)}\right)^2\left(\frac{\sqrt{\pi}\Gamma(\mu+1)}{2^{2\mu-1}\Gamma(\mu+\frac{1}{2})}-1\right),   
\end{align}
where the label $\varsigma,{\rm ML}$ should emphasise that these are the fluctuations associated with the coherent states based on the Mittag-Leffler  functions. We find that these fluctuations have a more complicated structure than in the case of the generalised coherent states introduced in this work, but also merge into the fluctuations of the standard harmonic oscillator coherent states if we choose $\mu=1$  and use $\Gamma(\frac{3}{2})=\frac{\sqrt{\pi}}{2}$. However, for $\mu\not=1$ the fluctuations involve a contribution with zero power of $\hbar$ and therefore whether these fluctuations are small is not entirely determined by $\hbar$ but for the second term depends on the value of $E_0$ being related to the coherent state labels. This is a property which as far as the semiclassical properties of the coherent states $|\varsigma;\mu\rangle_{\rm ML}$ are concerned can become problematic if we aim at keeping fluctuations small in general.

We have already seen that the semiclassical states introduced in \cite{2009:Laskin2} based on the fractional Poisson distribution presented in \cite{2003:Laskin1} for $\mu\not=1$ do not satisfy a resolution of identity, nor are they eigenstates of the annihilation operator. As the discussion above show they also do not yield the correct semiclassical limit for the Hamiltonian operator under consideration and further the size of the corresponding fluctuations can become large depending on the values of the classical labels of the coherent states. This leads to the  the conclusions that we would not consider these states as an appropriate set of semiclassical states for the operators of fractional power considered in this work. For the later purpose the generalised coherent states introduced in the article offer better functionality.

In order to obtain a better intuition on the physical interpretation of the states $|\alpha;\mu\rangle$, in the appendix in  section \ref{sectionComplexifierCoherentStatesForFractionalHeatKernel} we compare them to the standard complexifier coherent states. Furthermore, we very briefly discuss how the coherent states in \cite{2009:Laskin2} are related to the time fractional heat equation where the Mittag-Leffler function is involved in the construction of the corresponding heat kernels. 

The strategy to extend the range of polar coordinates to their covering space introduced in \cite{Klauder:1995yr} and still working with energy eigenstates of Schrödinger operators has been for instance also followed  in \cite{Sharatchandra:1997ha}. One of the motivations for such an extension comes in these works from the requirement of temporal stability. As we will not address this topic here and also in contrast to \cite{Klauder:1995yr,Sharatchandra:1997ha} still work with the (fractional) harmonic oscillator Hamiltonian we will discuss in the next subsection coherent sates in the context of polymerised Hamiltonians which might be the more suitable arena for the construction of these kind of coherent states that we are aiming at in this work. 
\subsection{Generalised Coherent States for fractional polymerised Hamiltonians}
\label{sec:PolyHam}
Although extending  polar coordinates to their covering space and considering the appropriate modified measure avoids the issue in the proof by Laskin in \cite{2009:Laskin2}, there seems to exist still some slight tension in working with the resolution of identity above for the following reason: When working with polar coordinates there exists a clear relation between the classical phase space labels $\alpha$ (or $q$ and $p$ respectively) and the range of integration and once this extension is made this relation is modified. At this stage this affects only the range for the classical labels, however these labels are associated with observables that become operators in the corresponding quantum theory. For this reason the above mentioned slight tension can be circumvented if we no longer work with the usual Schrödinger representation but consider the Hilbert space of quasi-periodic functions $L_2(\mathbb{R}_B,\d\mu_B)$, where $\mathbb{R}_B$ denotes the Bohr compactification of $\mathbb{R}$ and $\mu_B$ the Bohr measure. A consequence of this step is that in this representation we cannot just carry over the above proof because the harmonic oscillator Hamiltonian written in terms of the number operator cannot be implemented on that Hilbert space because the representation is only weakly continuous in either the position or momentum variable. On the one hand this requires a bit more work before a proof of the resolution of identity can be performed in this framework but fortunately there exits already various results in the literature on the polymerised harmonic oscillator \cite{Ashtekar:2002sn,Corichi:2007,BandStructurePolymer2013,Barbero:2014} that we can apply here. On the other hand it was not needed to use specific properties of the eigenstates of the usual harmonic oscillator in the proof above other than their completeness. As we will discuss below the required completeness is also given for the eigenstates of the polymerised harmonic oscillator, that are given in terms of periodic Mathieu functions,  in an appropriate superselection sector. For the reason that detailed results exists already in the literature we will be brief in this part and refer for more details to the references. 

We will follow closely the notation in \cite{Ashtekar:2002sn} although in the end we will use coherent states that differ from their shadow states.
The polymere Hilbert space is given by ${\cal H}_{\rm poly}:=L_2(\mathbb{R}_B,\mu_B)$. A generic element in ${\cal H}_{\rm poly}$ can be written as
$|\psi\rangle=\sum\limits_{x}\psi(x)|x\rangle$, where $\psi(x)$ is non-zero only at a countable set of points. The inner product in ${\cal H}_{\rm poly}$ reads $\langle x\, |\, x^\prime\rangle = \delta_{x,x'}$. The polymere representation of quantum mechanics chosen in \cite{Ashtekar:2002sn} is the one where the position operator $\hat{x}$ acts by multiplication and the translation operator $\hat{V}(\nu)$ is a 1-parameter unitary family that however fails to  be weakly continuous. Their action on ${\cal H}_{\rm poly}$ reads
\begin{equation*}
\hat{x} |x^\prime \rangle = x^\prime |x\rangle   \quad \hat{V}(\nu)|x^\prime \rangle = |x^\prime -\mu\rangle .
\end{equation*}
As a consequence and a difference to the Schrödinger representation is that the infinitesimal generator, that is a momentum operator, does not exist in this representation. The underlying algebra of the elementary operators, which can be understood as a polymere representation of the Heisenberg algebra, has the form
\begin{equation}
    [\hat{x},\hat{V}(\nu)]=-\mu\hat{V}(\nu)
\end{equation}
Therefore, the square of the momentum operator needs to be reexpressed in terms of the operators $\hat{V}(\nu)$. For this will choose the same form as in \cite{Ashtekar:2002sn} which is no unique choice but guided by the requirement that the spectrum of the polymerised  harmonic oscillator Hamiltonian should be non-degenerate. Similar to \cite{Ashtekar:2002sn} we introduce a fundamental length scale denoted by $\nu_0$ and assume that only those $\hat{V}(\nu)$ are relevant for which $\nu=N\nu_0$ for some integer $N$. Then the kinetic term in the harmonic oscillator Hamiltonian is replaced by $\widehat{K^2_{\nu_0}}$ given by
\begin{equation}
 \widehat{K^2_{\nu_0}} = \frac{1}{2\nu_0^2}\left( 2\mathds{1}-\hat{V}(\nu_0) - \hat{V}(-\nu_0)\right). 
\end{equation}
We introduce the length scale\footnote{This length scale has been denoted by $d$ in \cite{Ashtekar:2002sn} and furthermore our $\nu_0$ corresponds to their $\mu_0$ scale.}  $\ell:=\sqrt{\frac{\hbar}{m\omega_0}}$ that allows us to express the classical Hamiltonian already in terms of dimensionless position and momentum variables. Carrying this over to the quantum theory then the polymerised harmonic oscillator Hamiltonian has the form
\begin{equation*}
\hat{H}^{\rm poly} = -\frac{\hbar^2}{2m}\frac{1}{2\nu_0^2\ell^2}\left( 2\mathds{1}-\hat{V}(\nu_0) - \hat{V}(-\nu_0)\right)  +\frac{m\omega^2_0\ell^2}{2}\hat{x}^2    
\end{equation*}
The corresponding time-independent Schrödinger equation is a difference equation given by \cite{Ashtekar:2002sn}
\begin{equation}
\label{eq:SGLPolyx}
\Psi\left(x+\nu_{0}\right)+\Psi\left(x-\nu_{0}\right)-\left[2\mathds{1}-\frac{2 E}{\hbar \omega_0} \frac{\nu_0^{2}}{\ell^{2}}+\frac{x^{2}}{\ell^{2}} \frac{\nu_{0}^{2}}{\ell^{2}}\right] \Psi(x)=0
\end{equation}
Following \cite{Ashtekar:2002sn} we introduce the regular lattice $\alpha^{x_0}$ that is made of points $x_0+m\nu_0$ with $x_0\in [0,\nu_0)$. Then a suitable ansatz for the solution is given by
\begin{equation}
|\Psi_{x_{0}}\rangle=\sum_{m=-\infty}^{\infty} \Psi_{x_{0}}^{(m)} | x_{0}+m \nu_0\rangle
\end{equation}
and the Schrödinger equation becomes a recursion relation for $\Psi_{x_{0}}^{(m)}$ of the form 
\begin{equation}
\label{eq:SGLRecRel}
\Psi_{x_{0}}^{(m+1)}+\Psi_{x_{0}}^{(m-1)}-\left[2\mathds{1}-\frac{2 E}{\hbar \omega_0} \frac{\nu_0^{2}}{\ell^{2}}+\frac{\left(x_{0}+m \nu_0\right)^{2}}{\ell^{2}} \frac{\nu_0^{2}}{\ell^{2}}\right] \Psi_{x_{0}}^{(m)}=0.
\end{equation}
Now we take advantage of the fact that the non-separable Hilbert space ${\cal H}_{\rm poly}$ can be decomposed into a direct sum of separable ones labelled by $x_0$ according to
\begin{equation}
\mathcal{H}_{\text {Poly }}=\bigoplus_{\left.x_{0} \in \mid 0, \nu_0\right)} \mathcal{H}_{\text {Poly }}^{x_{0}}.
\end{equation}
The solution that one obtains from the recursion relation in \eqref{eq:SGLRecRel} is an element of  $\mathcal{H}_{\text {Poly }}^{x_{0}}$ and moreover the Hamiltonian acts in a way that it does not mix states between different $\mathcal{H}_{\text {Poly }}^{x_{0}}$. Hence, each of them is a superselection sector. As pointed out in \cite{Barbero:2014} as long as the polymerisation scale, that will be defined below more in detail is time-independent, which is the case here, we can restrict our analysis to one superselection sector only. We will chose the one for $x_0=0$. As explained in \cite{Ashtekar:2002sn,BandStructurePolymer2013} the eigenvectors in different superselection sectors can be associated with different boundary conditions of the solutions.  Let us denote the eigenstate in 
$\mathcal{H}_{\text {Poly }}^{0}$ by $|\psi_0\rangle$. To see that the Schrödinger equation in  $\mathcal{H}_{\text {Poly }}^{0}$ can be identified with the Mathieu equation we consider the momentum representation following \cite{Ashtekar:2002sn} and use
\begin{equation*}
\psi_0(k):=\langle k\,|\,\Psi_{0}\rangle=\sum_{m=-\infty}^{\infty} \Psi_{0}^{(m)} e^{-i k m \nu_0},\quad 
k \in\left(-\frac{\pi}{\nu_0}, \frac{\pi}{\nu_0}\right)
\end{equation*}
and obviously $\psi_0(k)$ has period $2\pi/\nu_0$. In the momentum representation the Schrödinger equation reads
\begin{equation}
\label{eq:DeqnHOPoly}
\frac{d^{2} \psi_{0}(k)}{d k^{2}}+\left(\frac{2\ell^2 E}{\hbar \omega_0}-\frac{2\ell^{4}}{\nu_0^{2}}+\frac{2\ell^{4}}{\nu_0^{2}}\cos \left(k \nu_0\right)\right) \psi_{0}(k)=0.
\end{equation}
Introducing the new variable $\phi\in [0,\pi]$ and parameters $a,q$
\begin{equation*}
\phi :=  \frac{k\nu_0+\pi}{2},\quad a:= \frac{8\ell^2}{\nu_0^2} \frac{E}{\hbar\omega_0}-\frac{8\ell^4}{\nu_0^4} ,\quad q:=\frac{4\ell^4}{\nu_0^4}    
\end{equation*}
we can rewrite \eqref{eq:DeqnHOPoly} as
\begin{equation}
\label{eq:MathieuDeqn}
\frac{d^2\psi_0}{d\phi^2}(\phi)+(a-2 q \cos (2 \phi))\psi_0(\phi)=0,
\end{equation}
which is the standard Mathieu equation. For the reason that $\cos(2\phi)$ has period $\pi$ we are looking for solutions $\psi_0(\phi)$ that are  $\pi$-periodic. These are exactly the known Mathieu functions of integral order, which in general have either period $\pi$ or $2\pi$ and are also called elliptic cosine and sine functions denoted by ${\rm ce}_{n}(\phi;q)$ and ${\rm se}_{n+1}(\phi;q)$ respectively with $n\in\mathbbm{N}_0$. In general also for $x_0\not=0$ the solutions of the Mathieu equation are of Floquet type, that is they can be written as $\psi_{x_0}(\phi)=e^{-2i(\phi-\frac{\pi}{2})\frac{x_0}{\nu_0}}u(\phi)$ with $u$ being a function with period $\pi$. Here the solutions have been chosen to satisfy the boundary condition $\psi_{x_0}(\phi+\pi)=e^{-i2\pi\frac{x_0}{\nu_0}}\psi_{x_0}(\phi)$ similar to \cite{Ashtekar:2002sn} such that the original function in the variable $k$ is of Floquet type with a shift of $\frac{2\pi}{\nu_0}$ and for $x_0=0$ this correspond to solutions being $\pi$-periodic in the variable $\phi$. Taking into account that ${\rm ce}_{n}(\phi;q),{\rm se}_{n+1}(\phi;q)$ are $\pi$-periodic for even labels and $2\pi$-periodic for odd labels the set of normalised eigenstates we are looking for is given by
\begin{eqnarray}
\label{eq:MatheiuEigenstates}
\psi_{0,n}(\phi):=\langle \phi\,|\,\psi_{0,n}\rangle & =& 
\begin{cases}
\sqrt{\frac{2}{\pi}}{\rm ce}_{2n}(\phi;q) & {\rm even\,\, fctns}\quad n\in\mathbb{N}_0 \\
\sqrt{\frac{2}{\pi}}{\rm se}_{2n+2}(\phi;q) & {\rm odd\,\,\,\, fctns} \quad n\in\mathbb{N}_0,
\end{cases}
\end{eqnarray}
where we have included a suitable normalisation factor. Such eigenfunctions were for instance also considered in \cite{HOlattice1986a} where the harmonic oscillator on a lattice was discussed but no semiclassical analysis was performed. For fixed values of $q$ the corresponding eigenvalues  for ${\rm ce}_{2n}(\phi;q)$ are denoted by $a=a_{2n}(q)$ and  for ${\rm se}_{2n+1}(\phi;q)$ we denote them as  $a=b_{2n+2}(q)$.
These are also called characteristic numbers in the context of the Mathieu equation. They can no longer be determined analytically but as discussed below for the computation of semiclassical expectation values their asymptotic form for large values of $q$ will be sufficient. For each fixed value of $q$ the Mathieu equation is a Sturm-Liouville problem with in our case periodic boundary conditions of the form $\psi_0(0)=\psi_0(\pi)$ and  $\psi^\prime_0(0)=\psi^\prime_0(\pi)$, where the latter follows directly from the periodicity of $\psi_0$ and the chain rule. Hence, we can apply the results known for Sturm-Liouville theory with periodic boundary conditions here and know that there exists a sequence of eigenvalues 
\begin{equation*}
a_0 < a_2 < a_4 < a_6 \cdots \to\infty\quad b_2 < b_4 < b_6  \cdots \to \infty    
\end{equation*}
with the following ordering among those two sets
\begin{equation*}
a_0 < b_2 < a_4 < b_4 < a_6 < b_6  \cdots \to\infty .  
\end{equation*}
As shown in \cite{McLachlanMathieu} these eigenvalues are non-degenerate for our case with $q\in\mathbb{R}$, $q>0$. Moreover 
$\{\sqrt{\frac{2}{\pi}}{\rm ce}_{2n}(\phi;q), \sqrt{\frac{2}{\pi}}{\rm se}_{2n+2}(\phi;q)\}_{n=0}^\infty$ provide an orthornomal basis for $L_2([0,\pi])$. Note that as usual ${\rm ce}_0$ is defined as ${\rm ce}_0(\phi;q)=\frac{1}{\sqrt{2}}$. To obtain a more compact notation we introduce the following notation for the eigenfunctions and eigenvalues denoted by $\lambda_{0,n}$
\begin{equation*}
\psi_{0,n}(\phi) =  \langle \phi\,|\,\psi_{0,n}\rangle 
\begin{cases}
\sqrt{\frac{2}{\pi}}{\rm ce}_{n}(\phi;q),\quad  \lambda_{0,n}(q):=a_n(q) & n\in\mathbb{N}_0\quad{\rm even} \\
\sqrt{\frac{2}{\pi}}{\rm se}_{n+1}(\phi;q), \quad \lambda_{0,n}(q):=b_{n+1}(q) & n\in\mathbb{N}_0\quad{\rm odd} 
\end{cases}
\end{equation*}
then we have 
\begin{eqnarray*}
\langle \psi_{0,n}\,|\,\psi_{0,m}\rangle &=& \delta_{n,m}\quad     
\sum\limits_{n=0}^\infty |\psi_{0,n}\rangle \langle \psi_{0,n}| =\mathds{1}_{\cal H}\quad {\cal H}=L_2([0,\pi])\\
\lambda_{0,0} <\lambda_{0,1} &<& \lambda_{0,2}<\cdots \quad \lim\limits_{n\to \infty} \lambda_{0,n} = \infty .
\end{eqnarray*}
For $q=0$ the eigenvalues $\lambda_{0,n}(q)$ reduce to $\lambda_{0,n}(0)=n^2$ and the eigenfunctions $\psi_{0,n}(\phi)$ either to $\cos(n\phi)$ or $\sin((n+1)\phi)$ depending on $n$ being even or odd. These completeness property allows us to use the eigenstates $|\psi_{0,n}\rangle$ to construct coherent state based on them in ${\cal H}_{\rm poly}^{0}$ using the strategy from \cite{Klauder:1995yr}. As in section \ref{sec:GenCS} we will introduce them with fractional labels and denote them in analogy as $|\alpha;0;\mu\rangle$ given by
\begin{equation}
\label{eq:GenCSPoly}
|\alpha ; 0 ;\mu\rangle:=\sum\limits_{n=0}^\infty \frac{(\alpha^\mu)^n}{\sqrt{n!}}e^{-\frac{1}{2}|\alpha|^{2\mu}}|\psi_{0,n}\rangle,
\end{equation}
 where the label $0$ refers to our choice $x_0=0$. The difference to the states $|\alpha;\mu\rangle$ in \eqref{eq:GenCS} is that $|\alpha;0;\mu\rangle$ here involve energy eigenstates of the polymerised Hamiltonian operator, whereas $|\alpha;\mu\rangle$ are constructed by means of ordinary harmonic oscillator eigenstates.
The normalisation of the states $|\alpha;0;\mu\rangle$ in \eqref{eq:GenCSPoly} follows from the normalisation of the eigenstates $|\psi_{0,n}\rangle$ and the proof for the  resolution of identity works analogously to the proof in section \ref{sec:GenCS} using the completeness of the states $|\psi_{0,n}\rangle$ in ${\cal H}_{\rm poly}^{0}$. Because there exist no annihilation and creation operators in ${\cal H}_{\rm poly}^{0}$  the coherent states $|\alpha;0;\mu\rangle$ cannot be understood as eigenstates of the annihilation operator here. However, this is not of similar importance here because the Hamiltonian operator can no longer be expressed in terms of a polynomial of annihilation and creation operators. As stated already in the last section the states considered in this work can be applied to fractional Hamiltonians but will not be the best states when fractional powers of position and momentum operator are considered.
Note that in contrast to \cite{Ashtekar:2002sn} we consider different coherent states even for the label $\mu=1$ because the shadow states used in \cite{Ashtekar:2002sn} are related to the complexifier coherent states \cite{GCS1,GCS2,GCS3,GCS4}, whereas the ones in \eqref{eq:GenCSPoly} are constructed in terms of the eigenfunctions of the polymerised Hamiltonian following \cite{Klauder:1995yr}.

In the following part of this section we will as in section \ref{sec:GenCS} use the harmonic oscillator Hamiltonian as a substitute for the fractional Hamiltonian and compute semiclassical expectation values with respect to the coherent states 
$|\alpha;0;\mu\rangle$. Considering the relation between $a$ and $E$ in the Mathieu equation \eqref{eq:MathieuDeqn} we realise that $\lambda_n(q)$ does not yield the energy $E$ directly but the relation reads
\begin{equation*}
\lambda_n(q)=\frac{8\ell^2}{\nu_0^2}\frac{E}{\hbar\omega_0}-\frac{8\ell^4}{\nu_0^4}=4\sqrt{q}\frac{E}{\hbar\omega_0}-2q,\quad q:=\frac{4\ell^4}{\nu_0^2}.
\end{equation*}
Hence, the polymerised Hamiltonian that enters the time-independent Schrödinger equation is given by
\begin{equation}
\label{eq:EffHPoly}
{\hat{H}}^{\rm poly} 
= \frac{\hbar\omega_0}{4\sqrt{q}}\left(-\frac{d^2}{d\phi^2}+2 q \cos (2 \phi)+2q\mathds{1}\right).
\end{equation}
The polymerised harmonic oscillator involves a discretisation scale $\nu_0$ that we expect to be smaller than the characteristic length scale $\ell$ of the harmonic oscillator. Thus, the values of $q$ that we are interested in are large since $q\to\infty$ for $\nu_0\to0$. At this stage the term involving the unit matrix looks potentially problematic for large $q$ but as we will see below similar to \cite{Ashtekar:1997fb} this term is cancelled by a contribution from the asymptotic behaviour of $\lambda_n(q)$. When we use  ${\hat{H}}_{\rm poly}$ as a substitution for the fractional operator  $({\hat{H}}_{\rm poly})^\mu$ as in section \ref{sec:GenCS} we need to rescale it for dimensional reasons by the factor $(\hbar\omega_0)^{\mu-1}$, see \eqref{eq:SubHO}, thus we introduce
\begin{equation*}
\widetilde{\hat{H}}_{\rm poly} := (\hbar\omega_0)^{\mu-1}{\hat{H}}^{\rm poly} 
\end{equation*}
and compute its semiclassical expectation values with respect to the states $|\alpha;0;\mu\rangle$
\begin{eqnarray}
\langle \alpha;0;\mu\,|\, \widetilde{\hat{H}}_{\rm poly} \,|\, \alpha;0;\mu\rangle 
&=&
(\hbar\omega_0)^{\mu-1}e^{-|\alpha|^{2\mu}}\frac{\hbar\omega_0}{4\sqrt{q}}\sum\limits_{n=0}^\infty \frac{(|\alpha|^{2\mu})^n}{n!}\left(\lambda_{0,n}(q)+2q\right)\nonumber \\
&=&
(\hbar\omega_0)^{\mu}\frac{e^{-|\alpha|^{2\mu}}}{4\sqrt{q}}
\sum\limits_{n\in\mathbb{N},{\rm even}}\frac{(|\alpha|^{2\mu})^n}{n!}\left(a_n(q)+2q\right)\nonumber\\
&&+(\hbar\omega_0)^{\mu}\frac{e^{-|\alpha|^{2\mu}}}{4\sqrt{q}}
\sum\limits_{n\in\mathbb{N},{\rm odd}} \frac{(|\alpha|^{2\mu})^n}{n!}\left(b_{n+1}(q)+2q\right).
\end{eqnarray}
Since we are interested in the large $q$ asymptotics first we use that for large $q$ we have $b_{n+1}(q)\sim a_n(q)$  \cite{MeixnerMathieu}, which allows us to combine the two separated sums involving either even or odd $n$'s to a single sum involving $a_n(q)$ only. Secondly, although an analytic expression for $a_n(q)$ is not available an analytic form for its asymptotics in terms of (inverse) powers of $q$ for up to order $q^{-\frac{5}{2}}$ is known, see for instance \cite{MeixnerMathieu}, which is more than sufficient for our application here. Explicitly, this expansion has the form
\begin{eqnarray*}
a_n(q)&\sim&
-2 q+2 s \sqrt{q}-\frac{1}{8}\left(s^{2}+1\right)-\frac{1}{2^{7} \sqrt{q}}\left(s^{3}+3 s\right)-\frac{1}{2^{12} q}\left(5 s^{4}+34 s^{2}+9\right) \\
&&-\frac{1}{2^{12} q^{\frac{3}{2}}}\left(33 s^{5}+410 s^{3}+405 s\right)-\frac{1}{2^{20} q^{2}}\left(63 s^{6}+1260 s^{4}+2943 s^{2}+486\right) \\
&&-\frac{1}{2^{25} q^{\frac{5}{2}}}
\left(527 s^{7}+15617 s^{5}+69001 s^{3}+41607 s\right)+{\cal O}(q^{-\frac{7}{2}}),\quad s:=2n+1.
\end{eqnarray*}
Using the asymptotic expansion and including terms up to the lowest order correction term, we get
\begin{eqnarray}
\langle \alpha;0;\mu\,|\, \widetilde{\hat{H}}_{\rm poly} \,|\, \alpha;0;\mu\rangle 
&=&
(\hbar\omega_0)^{\mu}e^{-|\alpha|^{2\mu}}
\sum\limits_{n=0}^\infty \frac{(|\alpha|^{2\mu})^n}{n!}\left(\frac{1}{2}(2n+1)-\frac{1}{32\sqrt{q}}((2n+1)^2+1)\right)+{\cal O}(q^{-\frac{3}{2}})\nonumber\\
&=&
E_0^\mu\left[1-\frac{1}{8\sqrt{q}}\left(2+|\alpha|^{2\mu}\right)   \right]+\frac{(\hbar\omega_0)^\mu}{2}\left[1-\frac{1}{8\sqrt{q}}\right]+{\cal O}(q^{-\frac{3}{2}})\
\end{eqnarray}
Compared to the result for the Schrödinger harmonic oscillator in \eqref{eq:ExpGenCSH} we obtain additional terms that involve inverse powers of $q$. Since we are in a sector where $q$ is large these are tiny and will vanish in the limit where the discretisation scale vanishes, that is $\nu_0\to 0$ corresponding to $q\to\infty.$ Because the coherent states $|\alpha;0;\mu\rangle$ involve the polymere Hamiltonian eigenstates $|\psi_{0,n}\rangle$ the corrections we obtain can be directly linked to the differences in the spectra of the usual and polymere Hamiltonian. Furthermore, these states allow a straight forward computation of semiclassical expectation values of (functions) of the polymere Hamiltonian operator. For $\mu=1$ they are the analogue of the harmonic oscillator coherent states $|\alpha\rangle$ if one follows the formalism in \cite{Klauder:1995yr} for the construction of coherent states. 
Likewise to what happens in full loop quantum gravity, see for instance the discussion in \cite{Towards1,Towards2}, also here we observe that the limit in which we send $\hbar$ as well as the discretisation scale $\nu_0$ to zero at the same time becomes a non-trivial step for higher order corrections because we have an interplay between very tiny and large terms multiplied by each other. Here this manifests in the $|\alpha|^{2\mu}$-term involved in the $E_0^\mu$-corrections. 
For the square of the fluctuations $\Delta\widetilde{\hat{H}}_{\rm poly}$ in the states $|\alpha;0;\mu\rangle$ we obtain
\begin{eqnarray}
 (\Delta\widetilde{\hat{H}}_{\rm poly})^2_{|\alpha;0;\mu\rangle }
&=&E_0^\mu(\hbar\omega_0)^\mu  \\
&&
+\frac{1}{\sqrt{q}}\left(\frac{2}{8}\right)\left(E^{2\mu}_0(\frac{1}{4}+\frac{1}{2}|\alpha|^{2\mu}) - \frac{1}{4}E^\mu_0(\hbar\omega_0)^\mu +\frac{(\hbar\omega_0)^\mu}{8}\right)\nonumber\\
&& +
\frac{1}{q}\left(\frac{1}{8^2}\right)\left(E^{2\mu}_0(10+4|\alpha|^{2\mu})-10E^\mu_0(\hbar\omega_0)^\mu
\right)\nonumber
\end{eqnarray}
Up to contributions that involve inverse powers of $q$ also the fluctuations agree with the result in the Schrödinger case. Because for higher order contributions the eigenvalues for the polymerised Hamiltonian involve higher powers of $s=2n+1$, higher powers of $|\alpha|^{2\mu}$ will contribute, which comes with inverse powers of $\hbar$ but also higher inverse powers of $q$. Hence, it will again depend on the values of the discretisation scale how tiny these contributions are. Note that this is a difference to the states from \cite{2009:Laskin2} for which the discretisation scale is absent but the fluctuations involved a term of order $\hbar^0$. In \cite{Ashtekar:2002sn} the eigenstates $|\psi_{0,n}\rangle$ were compared with the shadow states of the Schrödinger harmonic oscillator eigenstates. It was shown that to a good approximation these states agree, In principle one could use these Schrödinger shadows also to construct coherent states and it would be interesting to see whether and how the deviations from the Schrödinger result look like compared to what we obtained here.  These results discussed in this section show that also for the case when the coherent states carry fractional labels there exist a limit in which the results of the polymere quantum theory agree with those obtained for the usual Schrödinger harmonic oscillator.
\subsection{Application of the AQG-algorithm to our toy model with a fractional harmonic oscillator Hamiltonian}
\label{sectionSemiclassPerturbationFractionalHamiltonians}
Let us briefly compare the results obtained in this work to those that we get when we apply the AQG-algorithm \cite{Giesel:2006um} for computing semiclassical expectation values of fractional powers of the harmonic oscillator Hamiltonian. The idea of the AQG algorithm motivated by the non-polynomial form of the volume operator in LQG and the fact that is spectrum is not yet known, is the following: on the one hand one replaces the fractional volume operator by a power series of of operators that involve only integer powers and constructs the AQG-algorithm on the other hand in a way that the semiclassical limit still contains the correct fractional power. Details on the construction of the algorithm as well the application to the volume operator in LQG can be found in \cite{Giesel:2006um}. Here we only discuss the application to our quantum mechanical toy model. Given our fractional Hamiltonian $\hH_{\rm HO}^\mu$, then we are interested in computing the semiclassical expectation value
$\langle \alpha\,| \hH_{\rm HO}^{\mu}\,| \alpha \rangle$. To match with the notation in \cite{Giesel:2006um} let us define $\tilde{\mu}:=\frac{\mu}{2}$. In order to compute this expectation value we rewrite $\hH_{\rm HO}^{\mu}=(\hH^2_{\rm HO})^{\tilde{\mu}}$ and define $\hat{Q}:=\hH_{\rm HO}$, which is polynomial in the elementary operators. 
So for instance if we start with $\sqrt{\hH_{\rm HO}}$ then $\tilde{\mu}=\frac{1}{4}$ and we have $\sqrt{\hH_{\rm HO}}={}^4\sqrt{\hat{Q}^2}$. Then for $0<\tilde{\mu}\leq \frac{1}{4}$ it was shown that in order to compute $\langle \alpha\,| \hH_{\rm HO}^{\mu}\,| \alpha \rangle$ we can replace $\hH_{\rm HO}^{\mu}$ inside the semiclassical expectation value by the following operator valued power series
\begin{align}
\label{eqEstimateVolumeOpSemiclassicalPT}
    |\langle \alpha \,| \hat{Q}\,| \alpha \rangle|^{2\tilde{\mu}} \ls 1 + \sum\limits_{n=1}^{2k+1} (-1)^{n+1} \frac{\tilde{\mu} (1-\tilde{\mu}) \ldots (n-1-\tilde{\mu})}{n!} \lr \frac{\hat{Q}^2}{\langle \alpha \, |\hat{Q}\,| \alpha \rangle^2} - 1 \rr^n \rs.
\end{align}
The error that one makes by this substitution yields to corrections of order $\hbar^{k+1}$ and is thus smaller than the contribution of the highest order that one considers in this expansion. Note that in \cite{Giesel:2006um} this was analysed for the volume operator in LQC with SU($2$) complexifier coherent states. However, our toy model is simple enough that one can easily check that the assumptions that justify this expansion and substitution are satisfied here.  Considering that
 $\hH_{\rm{HO}} = \hbar \omega_0 \lr \had \ha + \frac{1}{2} \uop \rr$ and
\begin{align}
    \langle\alpha\, | \hH_{\rm{HO}} \, | \alpha\rangle = \hbar \omega_0 \lr |\alpha|^2 + \frac{1}{2}\rr  =\frac{m\omega_0^2q^2_0}{2}+\frac{p_0^2}{2m}+\frac{\hbar\omega_0}{2}=E_0+\frac{\hbar\omega_0}{2},
\end{align}
where we used that $\alpha=\sqrt{\frac{m\omega_0}{2\hbar}}q_0+i\sqrt{\frac{1}{2\hbar m\omega_0}}p_0$, up to the expansion of order $2k+1$ in $\hbar$, we can replace $\hH_{\rm{HO}}^{\mu}$ at the operator level by
\begin{align}
\label{eqEstimateFractionalHamiltonianSemiclassicalPT}
 \hH_{\rm HO}^\mu &= (\hH^2_{\rm HO})^{\tilde{\mu}} =   |\langle \alpha \, | \hH_{\rm{HO}}\, | \alpha \rangle|^{2\tilde{\mu}} \ls 1 + \sum\limits_{n=1}^{2k+1} (-1)^{n+1} \frac{\tilde{\mu} (1-\tilde{\mu}) \ldots (n-1-\tilde{\mu})}{n!} \lr \frac{\hH_{\rm HO}^2}{\langle \alpha\, | \hH_{\rm HO}\, | \alpha \rangle^2} - \uop \rr^n \rs\nonumber \\
    &=
    (E_0+\frac{\hbar\omega_0}{2})^{2\tilde{\mu}} \ls 1 + \sum\limits_{n=1}^{2k+1} (-1)^{n+1} \frac{\tilde{\mu} (1-\tilde{\mu}) \ldots (n-1-\tilde{\mu})}{n!} \lr \frac{\hH_{\rm HO}^2}{(E_0+\frac{\hbar\omega_0}{2})^2} - \uop \rr^n \rs,
\end{align}
in case we compute expectation values with respect to the standard harmonic oscillator coherent states $| \alpha \rangle$. This shows that to lowest order in $\hbar$ we can replace $\langle\alpha\, | \hH_{\rm{HO}}^{\mu} \, |\alpha\rangle$ by $\langle \alpha \, | \hH_{\rm{HO}}\, | \alpha \rangle^{\mu}$. 

A crucial ingredient in order to be able to define the operator valued power series expansion at the first place, is that the expectation value of $\langle \alpha \, | \hH_{\rm{HO}}\, | \alpha \rangle$ with respect to the coherent states $| \alpha \rangle$ can be computed. In our toy model this is obviously given but can become an issue in more complicated situations. Moreover, since the expansion involves inverse powers of $\langle \alpha \, | \hH_{\rm{HO}}\, | \alpha \rangle$ it can only be applied for those classical labels which yield non-zero expectation values and expectations values that are large enough such that corrections to the classical value stay small enough. If we compare the strategy to compute semiclassical expectation values in this work to the AQG algorithm we realise that the fractional power of the operators is treated differently.  In section \ref{sec:GenCS} we heavily rely on the fact that we start with a deparametrised constraint which enables us to shift the fractional power of the Hamiltonian to the temporal momentum, which for instance would not be available for the volume operator considered in \cite{Giesel:2006um}. Secondly, whereas for the AQG-algorithm one expands the fractional operator in terms of the integer powers, for the work in section \ref{sec:ConstCS} since the fractional power is attached to the temporal momentum operator $\hat{p}_t$, expectation values can be exactly calculated analytically in terms of Kummer's functions and no approximation scheme is necessary. As a consequence the correction to the classical value come with different powers in $\hbar$ since by construction for the AQG-algorithm from linear order on in the power expansion these have some integer power. On the other hand, for the AQG-algorithm no restrictions on the coherent state labels are assumed and thus such an expansion can be used on the kinematical as well as the physical level. The differences between the AQG-algorithm and the strategy we follow in section \ref{sec:GenCS} is that in the latter we modify the set of the coherent states, whereas the AQG-algorithms considers the standard harmonic oscillator coherent states. This modification allows to work with the linear power of the operator only instead of using an operator valued power series expansion.

\section{Summary and Conclusions}
\label{sec:SumConcl}
In this article we discussed possibilities to handle systems described by fractional powers of known Hamiltonians, shortly denoted as fractional Hamiltonians, respectively fractional Hamiltonian operators. Throughout the article we restricted our discussion to the fractional harmonic oscillator Hamiltonian operator as a toy model and to investigate as a first step in this direction how far we can get. 

The first approach we analysed in section \ref{sec:ConstCS} took as a starting point a constraint in deparametrised form $C=p_t+H_{\rm HO}^\mu$ with a corresponding physical Hamiltonian of the form $H_{\rm HO}^\mu$. Then, we considered a canonical transformation on the extended phase space in the variables $(t,p_t)$ as a kind of a so-called Euler rescaling in subsection \ref{sectionEulerRescaling} that allowed us to rewrite the constraint in a form where a fractional power is no longer attached to $H_{\rm HO}$ but only to the temporal momentum $p_t$. This has the advantage that we could then show that the standard kinematical harmonic oscillator coherent states yield a good semiclassical approximation of the constraint operator by means of the technique of Kummer's functions introduced in \cite{Giesel:2020jkz}. Afterwards in section \ref{sectionPhysicalCoherentStatesForFractionalHamiltonians} we applied a group averaging procedure following \cite{Ashtekar:2005dm} for the constraint with fractional temporal momentum and obtained the resulting physical coherent states and the physical inner product for this toy model. If we, as in \cite{Ashworth:1996yv,Ashtekar:2005dm}, require that the physical coherent states are peaked on the classical constraint surface, we can relate the semiclassical expectation value of $\hat{H}_{\rm HO}$ with respect to physical coherent states to fractional powers of the classical energy involved in the classical constraint. Interestingly, compared to the standard harmonic oscillator coherent states, it is exactly the modification of the states that results from the group averaging procedure which leads to this property.  
For the case that an inverse function exists, which was the case in our simple toy model, we can relate fractional powers of this semiclassical expectation value to the classical energy, something that also happens for the AQG-algorithm. On the one hand this shows that the so obtained physical coherent states have by construction some restriction on their labels which encodes dynamical properties of the system. However, on the other hand following this route in the final step an inverse function needs to be applied in order to get how the classical energy is related to the semiclassical expectation value of the Hamiltonian $H_{\rm HO}$.  The existence of this inverse function can become an issue if the $\hbar$ corrections of the linear power of the operator under consideration depend in a complicated way on the classical labels of the coherent states. A way out of this can be to change the set of coherent states and choose a set for which the $\hbar$ corrections take a simpler form and then this strategy of computing semiclassical expectation values can still be applied. Furthermore, we discuss in section \ref{sectionPhysicalCoherentStatesForFractionalHamiltonians} also how the results of the group averaging procedure and a reduced phase space quantisation of the same model are related and show that we obtain equivalent results in both cases.
Our results presented in this work extend the results of  \cite{Ashworth:1996yv,Ashtekar:2005dm} in the sense  that there only linear or quadratic powers of the elementary operators were analysed and here we considered fractional powers. We were able to extend their techniques to fractional powers by first shifting the fractional power from the 
Hamiltonian to the temporal momentum and second using the results in \cite{Giesel:2020jkz} that rely on the usage of Kummer's functions. 

In our second approach in section \ref{sec:GenCS} inspired by the coherent states based on a fractional Poisson distribution introduced in  \cite{2009:Laskin2} we analysed the question whether the labels of the coherent states can be adapted to Hamiltonians with fractional power. Although, the states in \cite{2009:Laskin2} yield fractional powers of the classical energy for the appropriately rescaled harmonic oscillator Hamiltonian, they do not satisfy a resolution of identity as originally claimed in \cite{2009:Laskin2}. We showed how the proof can be modified and adapted to our generalised coherent states constructed in subsection  \ref{sectionGeneralisedCoherentStatesForFractionalHamiltonians}. In contrast to the states in \cite{2009:Laskin2} the coherent states constructed in this article are still eigenstates of the standard annihilation operator. The reason why this is no longer the case for the Laskin states is that the exponential function usually involved in the standard harmonic oscillator coherent states is replaced by the so-called Mittag-Leffler function. Nevertheless, we can find a generalised annihilation operator which has the coherent state in \cite{2009:Laskin2} as an eigenstate. However, the algebra of these annihilation and creation operators does not reassemble the standard commutation relations and even depends on the number eigenstate. Moreover, in the semiclassical limit, that is the zeroth order of $\hbar$, the semiclassical expectation value yield not the expected classical result. This was one of the motivations for us to look for the generalised coherent states in section  \ref{sectionGeneralisedCoherentStatesForFractionalHamiltonians} which are still eigenstates of the annihilation operator but with an eigenvalue that involves fractional powers of the coherent states labels such as $\alpha
^\mu$ in our case. Since  by construction the fractional power is already involved in the eigenvalues and the labels and hence the construction of the coherent states, we then used the usual harmonic oscillator Hamiltonian as a kind of effective operator to substitute the fractional power Hamiltonian. As shown in this work this effective semiclassical computations yield good semiclassical properties. In contrast to the states in \cite{2009:Laskin2} they have the required classical limit. 
In addition we discuss the fluctuations of the states presented in  \ref{sectionGeneralisedCoherentStatesForFractionalHamiltonians} and the one from \cite{2009:Laskin2}. It turns out that   due to the  Mittag-Leffler  function involved in the latter their fluctuations have a more complicated structure. Problematic here is that these fluctuations also involve a term that is zeroth order in $\hbar$, which is not the case for the generalised states in \ref{sec:ReviewLaskinStates}.  As a consequence, the magnitude of these fluctuations is not mainly determined by $\hbar$ but  depends on the value of the classical energy $E_0$. Only in the specific case where the fractional label $\mu$ is set to $\mu=1$ this problematic term vanishes as expected because for $\mu=1$ these states agree with the standard harmonic oscillator coherent states. For a first brief intuition about these two sets of states we discuss in the appendix in section  \ref{sectionComplexifierCoherentStatesForFractionalHeatKernel} in which sense the states in \cite{2009:Laskin2} can  be understood as complexifier coherent states associated with the fractional heat kernel and their relation to the states in section \ref{sectionGeneralisedCoherentStatesForFractionalHamiltonians} in this context, where the fractional heat kernel was for instance discussed in \cite{GROTHAUS20151876,GROTHAUS20162732}. We do not elaborate this question in the appendix in full detail but just consider a specific limit of the Mittag-Leffler function in which such an analysis simplifies. 

For the reason that the extension of the range of the angular variable of the classical labels, originally suggested in \cite{Klauder:1995yr}, needed in the modification of the proof of the resolution of the identity of the generalised coherent states builds a bridge to the polymere framework, we also discussed the construction of analogue coherent states with fractional labels for the polymerised harmonic oscillator Hamiltonian in subsection \ref{sec:PolyHam}. We analysed their semiclassical properties and our results show that they also provide a suitable set of coherent states for the fractional Hamiltonian in the polymerised toy model. Furthermore, we obtain the expected behaviour that in the limit of a vanishing discretisation scale in the polymere framework the results agree with the corresponding Schrödinger case.

Finally, let us comment on the question whether the two approaches discussed here can be generalised to more complicated situation than the toy model considered in this work. For the group averaging approach as long as we restrict to deparametrised models even for more complex  Hamiltonian operators the constraints will be linearly in the temporal momentum, so the group averaging in the Hilbert space associated with the temporal degrees of freedom will have a similar effect. For instance in this work we considered coherent states based on the harmonic oscillator which can be also viewed as bosonic coherent states. There exist an extension to constrained fermionic systems introduced in \cite{Junker:1997qh}. We expect that for fermionic systems for which the dependence of the original Hamiltonian (without the fractional power) on the fermionic degrees of freedom is simple enough, the techniques of section \ref{sec:ConstCS} can be also carried over to those systems. However, in general the coherent states of the remaining degrees of freedom might not be so well adapted to the Hamiltonian as considered here and then the  relation to the classical energy might no longer be so easily obtained. Nevertheless, any suitable coherent states should have the property that in lowest order of $\hbar$ one obtains the classical energy plus small corrections and thus as far as only a few corrections next to the leading order are considered this can be applicable tool. For more general applications it will depend on the specific form of the Hamiltonian. For instance the quantum mechanical analogue of the Hamiltonian one considers in deparametrised models of General Relativity are of the form $\hat{H}=\left(f_1(\hat{q})\hat{p}^{\mu_1}f_2(\hat{q})\right)^{\mu_2}$, where $\mu_1,\mu_2$ are fractional powers and $f_1,f_2$ are polynomial or exponential functions respectively. For the outer fractional power $\mu_2$ the techniques presented in section \ref{sec:ConstCS} and \ref{sec:GenCS} can be applicable in case the set of coherent states that ones uses also approximate the function inside the outer fractional power, that is $f_1(\hat{q})\hat{p}^{\mu_1}f_2(\hat{q})$,  semiclassically sufficiently well.  For the inner fractional power $\mu_1$ the strategy in section \ref{sec:ConstCS} is not applicable. Here techniques like the AQG-algorithm \cite{Giesel:2006um}, the usage of Kummer's functions \cite{Giesel:2020jkz} or a choice of a different set of coherent states better adapted to the fractional operator than the standard harmonic oscillator ones along the lines of the discussion in section \ref{sec:GenCS} will be preferred.
 As far as our second approach in section \ref{sec:GenCS} is considered that works with coherent states involving fractional labels further more complicated applications need to be considered in order to understand their utility in full detail. We expect that these states can be useful for observables that are constructed from fractional powers of $\alpha$ and its complex conjugate as analysed in this work. If we consider instead observables that involve fractional powers of $q$ and $p$ instead we guess that the method of using Kummer's functions in \cite{Giesel:2020jkz} are favoured, see also our discussion regarding this point in the appendix \ref{sectionComplexifierCoherentStatesForFractionalHeatKernel}. For more insights and a better understanding this needs to be investigated in future applications. 
\section*{Acknowledgements}
A. Vetter thanks the Heinrich-Böll foundation for financial support at an early stage of this project.
We thank Andrea Dapor for many helpful discussions.
\begin{appendix}
\section{Complexifier coherent states and the time fractional heat kernel}
\label{sectionComplexifierCoherentStatesForFractionalHeatKernel}
In the complexifier approach \cite{GCS1,GCS2,GCS3,GCS4} coherent states are constructed as analytic continuations of the heat kernel, yielding directly the coherent states of (\ref{eq:GenCS}) for the choice $\mu=1$ and hence the standard harmonic oscillator coherent states. For an introduction to complexifer coherent states, see for instance \cite{GCS1}. Given the coherent states introduced in \cite{2009:Laskin2}, whose properties are summarised in section \ref{sec:ReviewLaskinStates}, we address the question in which sense this set of coherent states can be understood as complexifier coherent states for the fractional heat kernel shown in (\ref{eq:FractionalHeat1}) below. For simplicity in the following part we assume that $x,t,\hbar$ have been chosen to be dimensionless, which can always be achieved in units where we set the speed of light $c=1$. To obtain a fractional heat equation one replaces the temporal involved derivative in the heat equation by a so-called Caputo fractional derivative of order $\mu$ with $0<\mu\leq 1$ denoted by $\leftidx{^C}D^{\mu}_{0+}$ and the spatial derivative by  a fractional derivative of order $\nu$ given by $\leftidx{^C}D^{\nu}_{0+}$ with $0<\nu\leq2$. The equation for the fractional diffusion can then be written as
\begin{equation*}
\left(\leftidx{^C}D^{\mu}_{0+}u\right)(t,x)=\frac{1}{2}\left(\leftidx{^C}D^{\nu}_{0+}u\right)(x,t),\quad u(0,x)=f(x).    
\end{equation*}
Standard diffusion can be obtained by choosing $\mu=1, \nu=2$. Other prominent cases discussed in the literature, see for instance \cite{MAINARDI2005} and references therein, are the space fractional heat equation where $\mu=1$ and $0<\nu<2$, the one for neutral fractional diffusion with $0<\mu=\nu<2$ and the time fractional diffusion with the choices $\nu=2$ and $0<\mu<1$. For our discussion the latter one will be relevant and hence we only focus on that one from now on. Heat kernels for the time fractional heat equation have for instance been constructed in  \cite{Scheider:1989,Kochubei:1990,Mainardi:1995,eidelman2003cauchy}.
Using that the Caputo fractional derivative vanishes on constant functions as well as for continuous functions $u$ we have that $\left(\leftidx{^C}D^{\mu}_{0+}I^\mu_{0+}\,u\right)(t,x)=u(t,x)$, where $I^\mu_{0+}$ is a Riemann-Liouville integral. Then, the time fractional heat equation can be expressed as
\begin{eqnarray}
\label{eq:FractionalHeat1}
u(t,x)&=&u(0,x)+\frac{1}{2}\left(I^\mu_{0+}\frac{\partial^2}{\partial x^2}u(.,x)\right)(t) \nonumber \\    
&=&f(x)+\frac{1}{2\Gamma(\mu)}\int\limits_0^t ds (t-s)^{\mu-1}\frac{\partial^2}{\partial x^2}u(s,x).
\end{eqnarray}
In the context of fractional differential equations a useful property of the Mittag-Leffler function is the following
\begin{equation*}
\left( \leftidx{^C}D^{\mu}_{0+}E_\mu\right)(\lambda x^\mu)=\lambda E_\mu(\lambda x^\mu),\quad x\in\mathbbm{R}.   
\end{equation*}
Thus, the Mittag-Leffler function is the analogue of the exponential function for fractional differential operators since for $\mu=1$ the equation above just involves the first derivative of the exponential function. 
As discussed in \cite{GROTHAUS20151876,GROTHAUS20162732} a heat kernel for the fractional heat equation can be derived in the framework of grey Brownian motion and can be expressed in terms of the Mittag-Leffler function as
\begin{equation*}
\leftidx{^{\rm ML}}\rho_t(x,y;\mu)=\frac{1}{2\pi}\int\limits_{\mathbbm{R}} d\lambda e^{i\lambda(x-y)}E_{\mu}(-\frac{1}{2}\lambda^2 t^\mu)rm),    
\end{equation*}
which for $\mu=1$ can be integrated analytically and becomes the usual heat kernel $\rho_t(x,y)=\frac{1}{\sqrt{2\pi t}}exp(-\frac{1}{2t}(x-y)^2)$. The label ${\rm ML}$ here was chosen to emphasise that this heat kernel is based on the Mittag-Leffler function $E_\mu$.
To answer the question whether the coherent states in \cite{2009:Laskin2} can be understood as complexifier coherent states of the time fractional heat kernel, the analytic continuation of the integral involved in the heat kernel needs to be analysed in more detail which will be beyond the scope of this work here. 
However, since the asymptotics of the Mittag-Leffler function for large and small arguments is well known as a first step into this direction we can analyse this question in the limit where $t$ tends to zero that corresponds for the complexifier coherent states to the limit in which $\hbar$ is sent to zero and hence the semiclassical limit. Following \cite{Atkinson:2011} we use the series expansion for the Mittag-Leffler function given by
\begin{equation*}
E_\mu(z)=\sum\limits_{n=0}^\infty \frac{z^n}{\Gamma(1+\mu n)}\simeq 1+\frac{z}{\Gamma(1+\mu)}+o(|z|^2),   
\end{equation*}
where we neglected all terms higher than linear order in $z$ since we are interested in the asymptotic form  for small $|z|$. The right hand side can be read as the linearisation of a stretched exponential of the for $\exp(z/\Gamma(1+\mu))$. If we use this approximation in the heat kernel $\leftidx{^{\rm ML}}\rho_t(x,y;\mu)$, then we can perform the integral over $\lambda$ and obtain for $|z|\ll 1$
\begin{equation}
\label{eq:ApproxHKML}
\leftidx{^{\rm ML}}\rho_t(x,y;\mu) \simeq \frac{1}{\sqrt{2\pi t^\mu/\Gamma(1+\mu)}}
e^{-\frac{\Gamma(1+\mu)}{2t^\mu}(x-y)^2}. 
\end{equation}

As far as the generalised coherent states in (\ref{eq:GenCS}) are considered their non-normalised form can be formally obtained as complexifier coherent states from the standard heat kernel with a generalised map of the form $t\to\ell^2=\frac{\hbar}{m\omega_0}$ and $y\to\ell(\sqrt{2})^\mu\alpha^\mu$, where $\ell$ is included for dimensional reasons such that $[x]=[y]$ and the argument of the exponential function is dimensionless. This yields 
\begin{equation}
\label{eq:AppPsiCompl}
\Psi_{q_0,p_0;\mu}(x)=\left[\rho_{\ell^2}(x,y)\right]_{y\to\ell(\sqrt{2})^\mu\alpha^{\mu}}=\frac{1}{\sqrt{2\pi\ell^2}}e^{-\frac{(x-\ell^{-\mu+1}(q_0+\frac{i\ell^2}{\hbar} p_0)^\mu)^2}{2\ell^2}},   
\end{equation}
The scaling  by the factor $\sqrt{2}$ was introduced just for later convenience. Note that in case that we work with dimensionless quantities and the special case $\mu=1$, we have the usual definition of the non-normalised complexifier coherent states $\Psi_{q_0,p_0}(x)=[\rho_\hbar(x,y)]_{y\to\sqrt{2}{\alpha}}$ used in \cite{GCS1}. Note the relation between the complexifier coherent states and the states $|\alpha;\mu\rangle$ can be understood as follows: we can rewrite $|\alpha;\mu\rangle=e^{\alpha^\mu\hat{a}^\dagger}|0\rangle$. Then using that in the position representation $|0\rangle$ is a Gaussian, expressing $\hat{a}^\dagger$ in terms of $\hat{q}$ and $\hat{p}$,  applying the Baker-Campbell-Hausdorff formula, it is easy to see that $\langle x|\alpha;\mu\rangle$ agrees with $\Psi_{q_0,p_0;\mu}(x)$ up to a phase. This requires to define  $\Psi_{q_0,p_0;\mu}(x)$ also in terms of dimensionless quantities that is $\Psi_{q_0,p_0;\mu}(x)=\frac{1}{\sqrt{2\pi\hbar}}e^{-\frac{(x-\alpha^\mu)^2}{2\hbar}}$, where we choose units in which $x,y$ and $\hbar$ are dimensionless, as also chosen in \cite{GCS1}. We reintroduced the dimensions for $x$ and $y$ in the equation \eqref{eq:AppPsiCompl} again to make the following discussion more transparent.

Considering this kind of generalised map in \eqref{eq:AppPsiCompl}  we have a sort of imbalance between the variable $x$ and the map for the variable $y$ because they do not have the same power except for $\mu=1$. However, a generalised map of the form $y\to \ell(\sqrt{2})^{\mu}\alpha^\mu$ is a convenient choice here because it fits well to the fractional power of the Hamiltonian operator. A consequence of using such a map is that expectation values of the position and momentum operator with respect to these coherent states will only be peaked around $q_0$ and $p_0$ respectively if we choose $\mu=1$. Let us consider the remaining cases for the position operator. Obviously in these cases it will in general not be peaked at fractional powers of $q_0$ since  $\ell(\sqrt{2})^\mu\alpha^\mu={\rm Re}(\ell(\sqrt{2})^\mu\alpha^\mu)+i{\rm Im}(\ell(\sqrt{2})^\mu\alpha^\mu)$ with ${\rm Re}(\ell(\sqrt{2})^\mu\alpha^\mu)=\ell^{-\mu+1}|r|^\mu\cos(\mu\phi)$ and ${\rm Im}(\ell(\sqrt{2})^\mu\alpha^\mu)=\ell^{-\mu+1}|r|^\mu\sin(\mu\phi)$ for $r:=\sqrt{q_0^2+\frac{\ell^4}{\hbar^2}p_0^2}$ and $\phi=\arctan(\frac{\ell^2 p_0}{\hbar q_0})$. Thus, for the again for dimensional reasons rescaled position operator $\ell^{\mu-1}\hat{x}$ the coherent state is peaked around $|r|^\mu\cos(\mu\phi)$. Likewise working in the momentum representation the appropriately rescaled momentum operator will be peaked at $\frac{\hbar}{\ell^2}|r|^\mu\sin(\mu\phi)$. However, as shown in \cite{Giesel:2020jkz} for fractional powers of the momentum operator the standard coherent states for $\mu=1$ are already sufficient to yield a good semiclassical approximation if one uses Kummer functions and their Fourier transform. The results presented in \cite{Giesel:2020jkz} carry over to fractional powers of the position operator by employing the Fourier transform. Hence, one would rather use that techniques for fractional position and momentum operators. Therefore, it is not of disadvantage that, these states for $\mu\not=1$ are not peaked around the fractional powers of $q_0$ and $p_0$ in case we want to compute semiclassical expectation values of $\hat{q}^\mu$ and $\hat{p}^\mu$ because for these two cases one can work directly with the fractional operators themselves.

Comparing the standard heat kernel $\rho_t(x,y)$ involved in the construction of the generalised coherent states with the states  based on $\leftidx{^{\rm ML}}\rho_t(x,y;\mu)$ these two states differ in the semiclassical limit by the stretching due to the $\Gamma(1+\mu)$ and this we could also see in the corresponding semiclassical expectation values. 
The same can also be seen if we absorb the $\Gamma(1+\mu)$ into a redefinition of the diffusion constant, then the heat kernel in (\ref{eq:ApproxHKML}) can be understood as the heat kernel associated with fractional Brownian motion, see for instance \cite{GROTHAUS20162732}, which is described by the following fractional diffusion equation
\begin{equation}
\frac{\partial}{\partial t^\mu}u(t,x)=k_\mu\frac{\partial^2}{\partial x^2}u(x,t),
\end{equation}
where the diffusion constant was chosen again to be $k_\mu=\frac{1}{2\Gamma(\mu+1)}$ but now it has dimension $[x]^2[t]^{-\mu}$ and this differential equation can be regarded as a heat equation with stretched time variable. We can easily rewrite this equation in a form more closely to the standard heat equation given by
\begin{equation}
\frac{\partial}{\partial t}u(t,x)=k_\mu\mu t^{\mu-1}\frac{\partial^2}{\partial x^2}u(x,t)
\end{equation}
and  for $\mu=1$ we have $\Gamma(2)=1$ and as expected we recover the standard heat equation.
\end{appendix}
\bibliography{mybib}{}
\bibliographystyle{unsrt}
\end{document}